\documentclass[prl,superscriptaddress]{revtex4}

\usepackage{graphicx,t1enc}
\usepackage{epsfig}
\usepackage{epstopdf}
\usepackage{amssymb,amsmath}
\usepackage{color}
\usepackage{amstext}
\usepackage{amssymb}
\usepackage{graphicx}
\usepackage{esint}

\definecolor{blue}{rgb}{0,0,1}

\begin{document}

\title[]{Pedestrian evacuations with imitation of cooperative behavior}

\author{Amir Zablotsky}\thanks{Present address: Universit\'e Grenoble Alpes, CNRS, Laboratoire
Interdisciplinaire de Physique, 38000 Grenoble, France.}
\affiliation{Instituto Balseiro, Bustillo 9500, (8400) Bariloche, Argentina.}

\author{Marcelo Kuperman}
\affiliation{Instituto Balseiro, Bustillo 9500, (8400) Bariloche, Argentina.}
\affiliation{Consejo Nacional de Investigaciones Cient\'{\i}ficas y T\'ecnicas, Argentina}
\affiliation{Gerencia de F\'{\i}sica, Centro At\'omico Bariloche (CNEA), (8400) Bariloche, Argentina.}

\author{Sebasti\'an Bouzat}\thanks{email: bouzat@cab.cnea.gov.ar.}
\affiliation{Consejo Nacional de Investigaciones Cient\'{\i}ficas y T\'ecnicas, Argentina}
\affiliation{Gerencia de F\'{\i}sica, Centro At\'omico Bariloche (CNEA), (8400) Bariloche, Argentina.}

\begin{abstract}

We analyze the dynamics of room evacuation for mixed populations that include both competitive and
cooperative individuals through numerical simulations using the social force model. Cooperative agents represent
well-trained individuals who know how to behave in order to reduce risks within high-density crowds. We con-
sider that competitive agents can imitate cooperative behavior when they are in close proximity to cooperators.
We study the effects of the imitation of cooperative behavior on the duration and safety of evacuations, analyzing
evacuation time and other quantities of interest for varying parameters such as the proportions of mixing, the
aspect ratio of the room, and the parameters characterizing individual behaviors. Our main findings reveal that
the addition of a relatively small number of cooperative agents into a crowd can reduce evacuation time and
the density near the exit door, making the evacuation faster and safer despite an increase in the total number
of agents. In particular, for long spaces such as corridors, a small number of added cooperative agents can
significantly facilitate the evacuation process. We compare our results with those of systems without imitation
and also study the general role of cooperation, providing further analysis for homogeneous populations. Our
main conclusions emphasize the potential relevance of training people how to behave in high-density crowds

\end{abstract}
\maketitle


\section{Introduction}

The study of crowd dynamics has recently gainered considerable attention, both in academic circles and practical applications. The challenge lies in understanding a complex system that requires the integration of physics and computational techniques, as well as social science principles related to decision-making processes and individual and collective behavior. Several studies \cite{helbing2005self,coscia2008first,moussaid2010walking,bellomo2013microscale,cristiani2014multiscale} have investigated this area from a fundamental science perspective. Meanwhile, in terms of practical applications, mathematical models of pedestrian dynamics and simulation software have become essential for evaluating safety conditions in building and public facility design \cite{Johansson2008crowd,bellomo2016human}. This is important not only for emergency situations, where preventing fatalities is paramount, but also for regular pedestrian flow and individual comfort.
 
One of the most critical issues in crowd dynamics is the evacuation of enclosures through small exits or bottlenecks, which poses a significant risk to people and has been responsible for several tragic events. To understand the fundamental aspects of evacuations, numerous controlled experiments have been conducted \cite{helbing2005self,huang2008static,heliovaara2012pedestrian,guo2012route,GARCIMARTINEXP,GARCIMARTIN,von2017empirical,NICOLAS201730,seyfried}. However, due to the inherent difficulties and risks associated with experiments, mathematical models are crucial in unveiling the dynamics of evacuation processes. Numerical simulations enable researchers to analyze a wide range of scenarios and conditions. Various modeling frameworks have been employed to analyze pedestrian dynamics, including fluid-dynamics models \cite{HELBFLUID}, cellular automata \cite{BOUZATGAMETHEORY,nicolascaut,DONGME}, agent-based models \cite{heliovaara2012counterflow,BATTY,victor}, and particle-like descriptions of pedestrians, such as the social force model \cite{helb1995,helb2000,FRANK20112135,cornes}. These models are vital in gaining a deeper understanding of evacuation dynamics and devising strategies to minimize the risks involved.

Models that assume the same behavior for all the individuals in a crowd can provide useful information for understanding the generalities of pedestrian motion and evacuations. However, an attempt to describe the collective dynamics more precisely requires the consideration of the heterogeneity of people's behavior, as this can lead to significant deviations from the results obtained in homogeneous models \cite{CAOSFM,nicolascaut,nicolas2018,CORNESPANIC}.  A simple way of modeling heterogeneous populations is to reduce the wide variety of possible behaviors to only two categories \cite{BOUZATGAMETHEORY,CHENG2018485,GUAN2020124865}. One category corresponds to individuals that can be refered to as {\em cooperative} \cite{CHENG2018485} or {\em patient} \cite{GUAN2020124865}, for instance, while the other category corresponds to pededestrians that may be called {\em competitive}, {\em selfish} or {\em impatient}. In general, a cooperative agent represents an individual who, despite being focused on leaving the enclosure, is aware of the presence of others, keeps calm, is not aggressive, and does not rush. In contrast, a competitive agent tends to move faster than normal and rush for empty spaces. It is important to note that different degrees of cooperativeness and competitiveness can be considered, depending on the specific scenario. For instance, competitive agents could simply correspond to hurried or disrespectful pedestrians in non-emergency situations, or they could represent people dominated by panic who push and run over others in an emergency. On the other hand, several studies have modeled the attitudes of the pedestrians considering a continuous variable instead of only two possible states, usually regarding panic contagion, both for cellular automata \cite{nicolascaut} and particle-like \cite{ZHENG,CAOSFM} descriptions.

The phenomenon of panic emergence and contagion within congested crowds is considered as one of the main causes of targedies in pedestrian evacuations \cite{helb2000,XU}. Panic tends to increase the velocity of pedestrians and can produce areas of high pressure leading to falls, injuries and fatalities. From the point of view of modeling, panic contagion can be thought as the propagation of competitive behavior and has been the subject of many studies (see for instance \cite{helb2000,nicolascaut, ELZIE, CORNESPANIC, Wang_2022} and references therein). On the other hand, some studies have analyzed the role of cooperative behaviors in evacuations \cite{CHENG2018485, GUAN2020124865, Ma} with limited focus o propagation. As general trends, several studies suggest that while highly competitive behaviors tend to increase the risks and duration of evacuations, cooperative or patient attitudes lead to the contrary effects. In particular, it is a well established fact that cooperative attitudes in homogeneous populations can reduce the evacuation time and risks \cite{helb2000, GARCIMARTINEXP, GARCIMARTIN,victor}.  

In this context, we use numerical simulations to investigate how the propagation of cooperative attitudes, if achieved, may facilitate the evacuation through a narrow exit of a population whose pedestrians are in a considerable hurry to exit. It is worth noting that, while the spreading of panic (or of a selfish attitude) is a natural and explosive process related to the instinctive response of escaping from danger, the propagation of cooperative attitudes (i.e. the keeping of calm and the imitation of gentle and patient attitudes) can involve rather anti-instinctive actions that may require training and awareness. Hence, when assuming that cooperative attitudes propagate, we will be implicitly assuming that the pedestrians have a significant (although not homogeneous) degree of education concerning the benefits of keeping calm and behaving in prescribed forms. As we will show, this assumption leads to notable advantages for the evacuation process in the simulations. Our ultimate goal is to highlight the relevance that an appropriate education on how to behave within high-density crowds may have. 

In our studies we consider the social force model to simulate the evacuations. In contrast to many agent based or cellular-automata models, the social force model explicitly includes the contact (mechanical) force between pedestrians, which is a desirable ingredient regarding the clogging dynamics through narrow exits. We consider a population of competitive pedestrians, to which we add a variable number of what we call cooperative agents. The competitive agents represent hurried individuals with parameters that may correspond to a situation close to the onset of panic \cite{helb2000, CORNESPANIC}. However, we assume that they have a minimal degree of education or awareness, making them susceptible of being calmed, at least temporally, by the cooperative agents. The latter, on the other hand, represent better-trained pedestrians with a deeper degree of understanding of the relevance of remaining calm within congested crowds and with some capacity to influence their neighbors. Then, we assume that the competitive agents turn to adopt a cooperative behavior when they are close enough to a cooperative agent. In different scenarios, the cooperative agents are assumed to have a more patient or, alternatively, a more cautious attitude than competitive ones, as we will explain later. We analyze various quantities that characterize the evacuation process and its safety as functions of the number of added agents and of other relevant parameters. As we will show, depending on the type of scenario and conditions, even a limited spread of cooperativeness and a relatively small number of cooperative agents can lead to a considerable easing of the evacuation and to an increase in safety.

It is worth remarking that our studies provide a numerical assessment of what could happen in populations with a degree of education that may need to be higher that the one we have in most societies nowadays, particularly in terms of imitation of cooperation. While there is evidence of the emergence and propagation of cooperative behaviors in emergency situations (see \cite{CHENG2018485} and references therein), our simulations highlight the potential impact of education and awareness in promoting cooperative behaviors during pedestrian evacuations. Particularly, if such an education could enable rational (even anti-instinctive) individual decisions to prevail over instinctive reactions known to cause catastrophic events. The results could eventually be contrasted to controlled experiments with actual pedestrians that should follow the line of the experiments in Ref. \cite{NICOLAS201730} but including protocols for imitation. Further interdisciplinary studies are needed before suggesting concrete decisions on education. Beyond our main objective, the present work provides new insights and results concerning the dynamics of the social force model, both for inhomogeneous and homogeneous populations, that may be of interest for basic studies and applications of the model as well.

The paper is organized as follows. In Section 2, we introduce the model for pedestrian dynamics and the rules for imitation. We also describe the quantities used to analyze the simulations. In Section 3, we present the results. We first discuss the evacuation of homogeneous populations, then the evacuation of mixed populations in rooms with variable geometries.  After that, we consider mixed populations that combine different types of cooperative agents, and finally, for the sake of completeness, we analyze systems without imitation. In Section 4, we summarize our conclusions.


\section{Models and simulations}

\subsection{The Social force model}

\begin{figure}[ht]
\includegraphics[width=0.9\columnwidth]{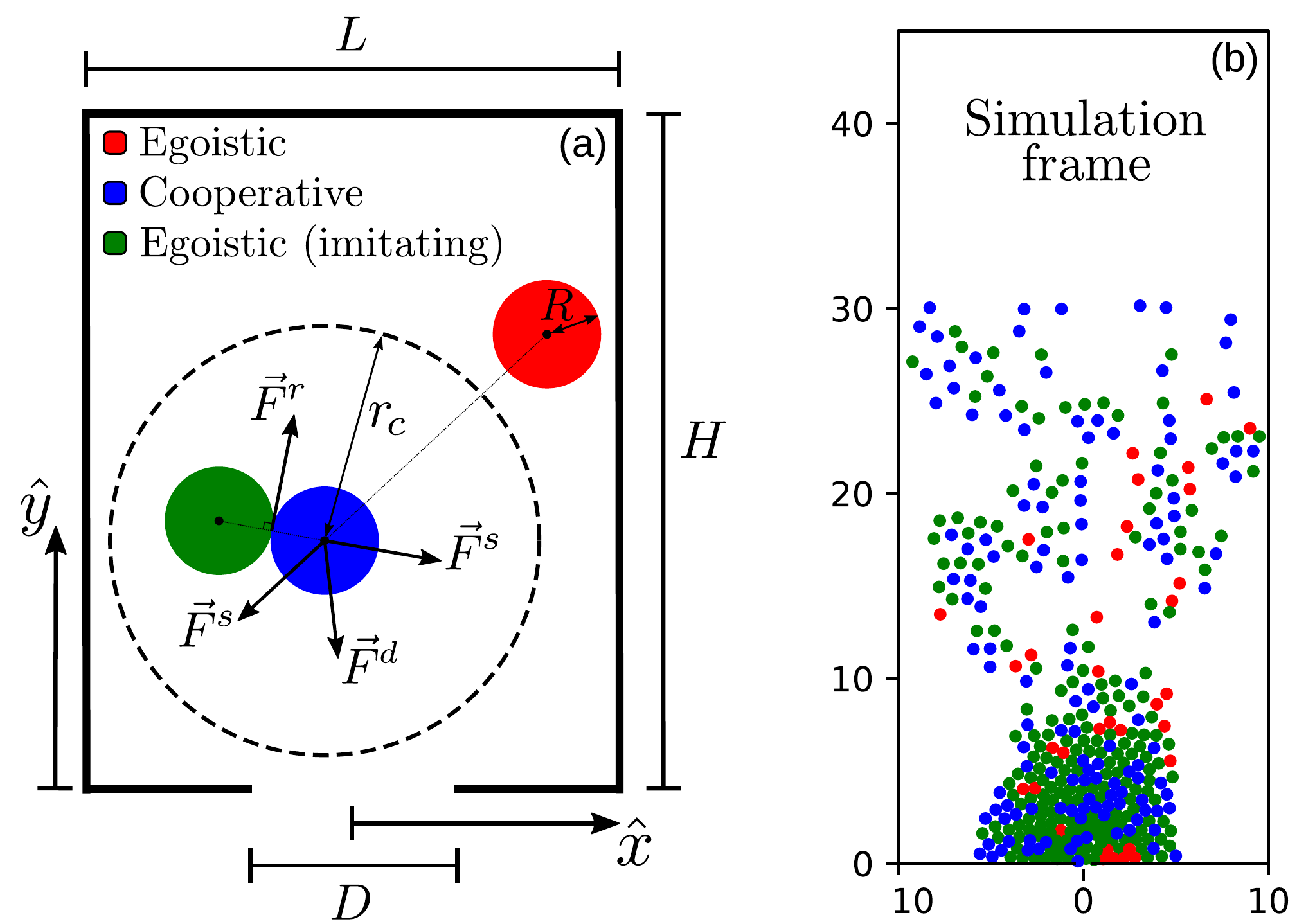}
\caption{\label{f-esquema} Diagram of the model. a) Rectangular room of size $L\times H$ with exit of size $D$. Inside the room, we show a  sketch of the interactions exerted on a cooperative agent (blue circle) by two competitive agents. The competitive agent that is inside the area of the radius $r_c$ imitates the cooperative behavior. For the sake of simplicity we do not illustrate the interaction with the walls which are defined in the text.  b) Simulation frame of an evacuation.}
\end{figure}

We analyze the evacuation of a rectangular room with sides of length $L$ and $H$, where $L\leq H$. The only exit has a size of $D$ and is located at the center of one of the sides with length $L$ (see diagram in Fig.\ref{f-esquema}.a). Unless indicated otherwise, we fix $D=1\,$m, which corresponds to two diameters of a pedestrian. Meanwhile, $L$ and $H$ are varied in different simulations.

To simulate the evacuation, we employ a standard version of the social force model \cite{helb2000,FRANK20112135,rozan,CORNES2023106218} with radially-symmetric repulsive social forces. The dynamical equations for a population of $N$ pedestrians with positions $\vec{r}_i$, where $i=1\dots N$, are given by
\begin{eqnarray}
\frac{d\vec{r}_i}{dt}&=&\vec{v}_i \nonumber \\
m_i\frac{d\vec{v}_i}{dt}&=&\vec{F}^d_i+\sum_{j\neq i}\vec{F}^s_{ij}+\sum_{j\neq i}\vec{F}^f_{ij}.
\label{sfm}
\end{eqnarray}

In this context, $\vec{v}_i$ represents the velocity of pedestrian $i$, and $m_i$ denotes their mass, which is set as $70{\rm kg}$ for all pedestrians in this study. Equation ($\ref{sfm}$) indicates that the motion of each pedestrian is influenced by three forces. First,
\begin{equation}
\label{Fdeseo}
\vec{F}^d_i=\frac{m_i\left(v_{d,i}\,\,\hat{e}_i-\vec{v}_i\right)}{\tau_i}
\end{equation}
is the self-propulsion or {\em desire} force of agent $i$, which is proportional to the difference between the 
actual velocity $\vec{v}_i$ and the desired velocity $v_{d,i} \hat{e}_i$. Here, $v_{d,i}$ is the desired speed and  
$\hat{e}_i$ is the unit vector indicating the desired direction of motion which is toward the nearest point within the exit.  
The parameter $\tau_i$ is the typical relaxation time that takes the agent to reach a fixed desired velocity in the absence 
of other forces. In this work, we set $\tau_i=0.5\,$s for all $i$ \cite{helb1995}. 

The second force term in Eq. (\ref{sfm}) corresponds to the repulsion {\em felt} by agent $i$ due to the presence of agent $j$, given by
\begin{equation}
\label{Fsocial}
\vec{F}^s_{ij}=A_ie^{\frac{\left(R_i+R_j\right)-\Vert\vec{r}_i-\vec{r}_j\Vert}{B_i}}\hat{n}_{ij}.
\end{equation}
Here, $R_i$ is the radius of agent $i$ and $\hat{n}_{ij}$ is the unit vector pointing from $\vec{r}_j$ to $\vec{r}_i$. The parameter $A_i$ defines the repulsion felt by agent $i$ when touching agent $j$  (i.e., when they are separated by a distance $R_i+R_j$), while $B_i$ defines the exponential decay of the force with the distance. 
For distances larger than $R_i+R_j$, $\vec{F}^s_{ij}$ represents the social force, while when the two agents overlap it  also accounts for the mechanical body force.  The definition given in Eq.(\ref{Fsocial}), that considers a single continuous function for describing both the social force and the physical force in the normal direction corresponds to the assumption  used in \cite{rozan,CORNES2023106218,STICCO202042,FRANK20112135}. Other versions of the social force model include a separate description of the physical force that may be relevant to reproduce particular experiments in detail. However, as the general phenomenology  observed in the simulations of both types of models is essentially the same and our studies do not involve fine tuning of the parameters, here we choose the simplest version. 
 The last term in Eq. (\ref{sfm}) corresponds to the tangential friction force exerted by agent $j$ on the agent $i$:
\begin{equation}
\vec{F}^f_{ij}=\kappa\, \Theta \left(\left[R_i+R_j\right]-\Vert\vec{r}_i-\vec{r}_j\Vert\right)\left(\left[R_i+R_j\right]-\Vert\vec{r}_i-\vec{r}_j\Vert\right)\left(\Delta\vec{v}_{ij}\cdot\hat{t}_{ij}\right)\,\hat{t}_{ij}. 
\end{equation}
As $\Theta$ represents the Heaviside function, this force acts only if the two agents are overlapped. 
It has the direction of the tangent versor $\hat{t}_{ij}=\left(-n_{ij}^{(2)},n_{ij}^{(1)}\right)$ and is proportional to the overlap and also to the tangent component of the difference of velocities  $\Delta \vec{v}_{ij}=\vec{v}_j-\vec{v}_i$ and to the friction constant $\kappa$ \cite{helb2000}. Following the proposals of previous works \cite{helb2000}, we set $\kappa =2,4\times10^{5}\,\text{kg}\,\text{m}^{-1}\,\text{s}^{-1}$ and we consider $B_i=0.08\,$m and $R_i=0.25\,$m for all $i$ \cite{FRANK20112135}.
In Fig. \ref{f-esquema}. we show a diagram of the room and of the forces acting on an agent marked as {\em cooperative}. 

Both the social force $\vec{F}^s$ and the friction force $\vec{F}^f$ in Ec. 1 consider not only the interaction between pedestrians, but also the interaction between the pedestrians and the walls. For agent $i$, the social force felt as a result of its proximity to each wall is described by Eq. 3, where $R_j=0\,$m and $\vec{r}_j$ denotes the point of the wall closest to agent $i$ \cite{helb2000}. Similarly, the friction force that the walls exert on agent $i$ is given by Eq. 4, where $R_j=0\,$m, $\vec{v}_j=0\,$m/s and $\vec{r}_j$ is the point of the wall closest to agent $i$ \cite{helb2000}.


Note that, according to Eq. \ref{Fdeseo} with the standard parameters considered \cite{helb1995}, the initial acceleration of an isolated pedestrian starting from rest may be significant high, of the order of the accelerations attained by high-performance athletes. However, this may only occur for a very short time window ($\sim 0.2$ s) after the initial condition of the simulation of a low density system. In any other situation during simulations of evacuations, agents either move with non vanishing velocities (of the order of $v_d$) or are not entirely isolated, thereby being subject to social and contact forces that decrease their accelerations. The simple linear model for the desired force is primarily intended to describe the relaxation of the velocity to the desired one for an agent within a crowd, rather than to reproduce in detail the whole process of acceleration of an isolated pedestrian.


The dynamical equations are solved using the velocity Verlet algorithm with a time-step of $1\,$ms \cite{vel-verlet}.
The initial positions of the agents, regardless of their type, are set at random within the room, taking care of leaving at least $0.5\,$m (one pedestrian's width) of free space between any two agents and between agents and walls. For each parameter set considered we perform $50$ realizations of the evacuation dynamics with different random initial conditions. 

\subsection{Mixed populations and imitation of cooperative behavior}

As stated in the introduction, our study will investigate the evacuation dynamics of mixed populations composed of cooperative and competitive agents, the latter also referred to as egotistic agents. In particular, we will explore the possibility of imitation of cooperative behavior by competitive agents. We will denote the number of egotistic agents as $N_e$, and the number of cooperative agents as $N_c$.

In the various examples that we analyze, we consider two different types or {\em versions} of cooperative agents. The first type corresponds to cooperative agents defined by considering a desired velocity ($v_{d,c}$) smaller than the one used for egotistic agents ($v_{d,e}$). Cooperation in this case can be associated with a {\em patient} attitude. Note that a smaller value of desired velocity corresponds to a less hurried attitude, that would contribute to ordering the flux and decreasing the effective pressure. According to Eq. \ref{Fdeseo}, in a situation of clogging in which the actual velocities of the agents are slower than the desired ones, the competitive pedestrians would tend to accelerate faster than the patient cooperative ones, and they would also exert a stronger pressure on the agents that are ahead of them. The second type of cooperative agents considered are referred to as {\em cautious} agents. These are defined by considering a value of the parameter $A_i$ larger than the one used for competitive agents. Hence, in this case we set $A_c>A_e$ while the remaining parameters are the same for both agents types. Note that the parameter $A_i$ is used for defining the social force in Eq. \ref{Fsocial}. A larger value of $A_i$ implies that the agent tries to maintain a greater distance from their neighbors, thereby avoiding physical contact and leaving free space for other agents to move.

The dynamics of imitation assumes that competitive agents turn to adopt cooperative behavior if they are close enough to cooperative agents.  In most of our studies, we simulate mixed populations with two types of agents. Namely, the competitive pedestrians and only one type of cooperative agents (either patient or cautious). In these cases, the implementation of the imitation dynamics is as follows: a critical distance for imitation denoted as $r_c$ is introduced. If the distance between a competitive pedestrian and a cooperative agent is smaller than $r_c$, the competitive agent adopts the parameters of the cooperative agent. However, if the distance becomes larger than $r_c$, the competitive agent reverts back to behaving as competitive. Therefore, imitation is not permanent, but rather instantaneously conditioned by the proximity to a cooperative agent. Moreover, only the original cooperative agents can be imitated, as competitive agents who are temporarily behaving cooperatively cannot be imitated. The imitation dynamics is the same for the cases in which we consider patient or cautious cooperative agents. In the first case, the competitive pedestrians change their value of $v_d$ when imitating cooperators, while in the second case they change $A_i$.

At the end of the work we simulate evacuations of mixed populations with three types of agents, i.e. the two types of cooperative agents in addition to the competitive ones. In this case, the dynamics of imitations considers the same imitation radius $r_c$, and is defined through a majority game algorithm that will be explained later.

The radius for imitation is set as $r_c=1\,$m throughout the work, as it is reasonable to expect that pedestrians within a distance of approximately $1\,$m may communicate through speaking or physical contact. For example, cooperative agents could try to calm other pedestrians by speaking, giving instructions or touching their shoulders. This parameter is kept fixed throughout our study, as the results are expected to be robust to small changes in $r_c$. We also investigate the limiting case where $r_c=0\,$m, which corresponds to evacuations without imitation of cooperative behavior. The other limiting case, $r_c=\max \{L,H\}$, would result in a fully-cooperative crowd with homogeneous behavior.

In the diagram of Fig. \ref{f-esquema}.a we illustrate the three types of agents and the distance for 
imitation. Fig. \ref{f-esquema}.b shows an instantaneous state of a simulation of a mixed population with competitive agents, cooperative agents, and imitators. Meanwhile, Table \ref{tabla} summarizes the model parameters used along the work.

 As mentioned in the introduction, the most relevant interpretation of the model of imitation is that the cooperative agents represent pedestrians with a higher degree of education and training on how to behave and calm other pedestrians compared to the competitive ones. The competitive pedestrians, on the other hand, represent hurried individuals which may be at the onset of panic \cite{helb2000,CORNESPANIC}, who have however 
some level of knowledge about the benefits of remaining calm and are thus susceptible to being influenced, at least temporarily, by the cooperative agents. By considering that only the originaly cooperative agents can be imitated we are implicitly assuming that competitive agents can be temporaly induced to remain calm or not to hurry but they cannot be induced to be able to calm other people. This can be compatible either with a lower degree of education, a lower conviction on the relevance of remaining calm, a lower leadership attitude or a higher degree of fear compared to cooperators. Note that competitive pedestrians are also unable to imitate the permanent temperance of the cooperative agents, as they begin to behave in a selfish way as long as they get far enough from cooperative agents. 

On the other hand, an alternative interpretation of the model is that cooperative pedestrians may represent security agents wearing recognizable uniforms instead of representing ordinary individuals of the population with a deeper understanding of the convenient behavior. Thus, they are clearly able to influence the behavior of their neighbors. 

It is important to note that, regardless of the interpretation considered for the cooperators, imitators are not the same as cooperators, and the conceptual significance of the model lies in the fact that the cooperative agents induce cooperativity in the competitive population, but the propagation is limited, as imitation is not permanent and imitators cannot be imitated. Our assumptions aim to demonstrate how a relatively modest degree of propagation of cooperativity can significantly improve evacuations. If these conditions were relaxed, for instance if competitive agents imitating cooperative behavior could also be imitated, then cooperativity would spread rapidly throughout the population, and most agents would start behaving like the cooperative agents. In such a case, the effects resulting from the propagation of cooperative behavior would be enhanced. 


It is worth noting that there are some differences between our assupmtions for the cooperative agents with those concerning {\em leader} agents found in the literature. It is commonly assumed that leader agents possess superior knowledge of the room's geometry, including the location of the exit, or have a heightened capacity for exploration, as evidenced in studies like \cite{HOU, LOPEZCARMONA, PELCHANO}. In contrast, our  cooperative agents do not have any additional spatial information compared to the competitive agents. Instead, they possess a deeper understanding of the benefits of remaining calm, moving slowly, and reassuring others during evacuation. On the other hand, the competitive agents are equally aware of the exit's position, but may lack the understanding of the advantages of patience and cautiousness demonstrated by the cooperative agents.

\begin{table}[]
\begin{tabular}{|c|cc|c|}
\hline
  \textbf{Parameter}         & \multicolumn{2}{c|}{\textbf{Values}}          & \textbf{Comment} \\ \hline
$\boldsymbol{R}$            &  \multicolumn{2}{c|}{$0.25\,$m}                                            &   Agent radius      \cite{helb2000, FRANK20112135}          \\ \hline
$\boldsymbol{m}$           & \multicolumn{2}{c|}{$70\,$kg}                                             &   Agent mass      \cite{FRANK20112135}         \\ \hline
$\boldsymbol{\tau_i}$       & \multicolumn{2}{c|}{$0.5\,$s}                                             &  Relaxation time to desired velocity    \cite{helb2000, FRANK20112135}         \\ \hline
$\boldsymbol{r_c}$           & \multicolumn{2}{c|}  {$0$ or $1\,$m}                  &  Critical distance for imitation, see text for details.            \\ \hline
$\boldsymbol{B}$              & \multicolumn{2}{c|}{$0.08\,$m}                                            & Decay length defining social force    \cite{helb2000, FRANK20112135}         \\ \hline
$\boldsymbol{\kappa}$       & \multicolumn{2}{c|}{$2.4\times10^5\,$kg$\,$m$^{-1}\,$s$^{-1}$}            &   Friction constant   \cite{helb2000, FRANK20112135}             \\ \hline
$\boldsymbol{A}_e$              & \multicolumn{2}{c|}{$2000\,$N}  &     Amplitude of social force for competitive pedestrians   \cite{helb2000, FRANK20112135}                    \\ \hline
$\boldsymbol{A}_c$              & \multicolumn{2}{c|} {$2000\,$N}  &   Amplitude of social force for {\em patient} cooperative  agents   \cite{helb2000, FRANK20112135}                    \\ \cline{2-4} 
              & \multicolumn{2}{c|} {Range $3000\,$N to $10000\,$N}  &   Amplitude of social force for {\em cautious} cooperative  agents   \cite{helb2000, FRANK20112135}                    \\ \hline
$\boldsymbol{v_{d,e}}$          & \multicolumn{2}{c|}{$3\,$m$\,$s$^{-1}$}         &   Desired velocity for competitive agents. See text for details.          \\ \hline
$\boldsymbol{v_{d,c}}$          & \multicolumn{2}{c|}{$3\,$m$\,$s$^{-1}$}          &   Desired velocity for {\em cautious} cooperative  agents.   See text for details.          \\ \cline{2-4} 
          & \multicolumn{2}{c|}{Range $0.25\,$m$\,$s$^{-1}$ to $2.25\,$m$\,$s$^{-1}$}          &   Desired velocity for {\em patient} cooperative  agents   See text for details.          \\ \hline
\end{tabular}
\caption{\label{tabla}  Model parameters considered along the work. The references indicated in column {\em Comment} correspond to articles where such values has been previously reported or considered. Note that for patient cooperative agents we set $A_c=A_e$ and vary $v_{d,c}$ in the indicated range (with $v_{d,c}<v_{d,e}$). Meanwhile, for cautious cooperative agents we set $v_{d,c}=v_{d,e}$ and vary $A_c$ in the indicated range (with $A_c>A_e$).}
\end{table}

\subsection{Quantities of interest}

Here we introduce the quantities of interest that we calculate along our work for the different sets of simulations. As stated in the introduction, we are mainly interested in analyzing how the presence of cooperative agents and the propagation of the cooperative behavior affect the duration and safety of the evacuations. The duration of the evacuations can be directly quantified by the evacuation time, meanwhile, we will characterize the safety of the evacuation by measuring the density around the exit as we explain below. Other relevant metrics to compute are the survival function, that characterizes the intervals between escapes, and the fundamental diagram relating density with velocities.

For each parameter set considered, we perform $50$ realizations of the dynamics with different random initial positions of the agents. Then, for each realization of the dynamics, the evacuation time $T$ is defined as the time it takes for $80\%$ of the pedestrians to leave the room \cite{sticco,rozan}. In our analysis, we focus on the median of $T$ computed over the $50$ realizations for each parameter set considered, and we also calculate the first and third quartiles to characterize the width of the distribution of $T$. For the sake of shortness, throughout the paper, the {\em evacuation time} $T$ (sometimes also called {\em exit time}) refers to the median of $T$ computed over $50$ realizations.

Another quantity of interest is the mean density around the door as a function of time, referred to as $\rho(t)$. For each time $t$, $\rho(t)$ is calculated as an average over realizations of the number of pedestrians located in a semicircle of radius $1$\,m around the center of the door, divided by the area of such semicircle. This is a measure of the crowding around the door and an indirect indication of the pressure felt by the pedestrians in that very zone. Relative small values of $\rho(t)$ characterize safe evacuations while large values of $\rho(t)$ indicate risks of injuries.  

We also analyze the survival function for the times between successive escapes denoted as $P\left(\Delta t > \tau\right)$. This function, which depends on the variable $\tau$, gives the probability that the time ($\Delta t$) between the escapes of two pedestrians who leave the room successively is greater than $\tau$ \cite{nicolascaut}. For each set of parameters, the survival function is computed from all the $\Delta t$ that occurred during the $50$ realizations of the evacuation.

Finally, we perform the calculation of the {\em fundamental diagram} \cite{funddiag}, which illustrates the relationship between the average velocity of the agents in the exit zone and the corresponding density. To obtain this diagram, we employ the same region surrounding the door that was used to compute $\rho(t)$. At every $100$ simulation steps, we determine the mean speed and density of the agents within the area, and then we average the results across multiple realizations. As previously stated, we compute the local density in the region as the number of agents inside it divided by its surface area \cite{SEYFRIED1,CAO1}.

\section{Results}

Our plan for the presentation of the results and analysis is as follows. First, in the next subsection we analyze the case of homogeneous populations. This is in order to fix some ideas on the role of the parameters $A$ and $v_d$ that we later use to define cooperative and competitive behaviors, and also to analyze the dependence on the geometry of the room. Then, we present our main studies of evacuations of mixed populations with imitation of cooperative behavior. We begin by analyzing the case of square rooms and, after that, we consider the case of long rooms. At the end of the section we present results for mixed systems without imitation, for the sake of comparison and completeness.

\subsection{Homogeneous populations: the faster-is-slower effect.}

\begin{figure}[ht]
\includegraphics[width=0.85\columnwidth]{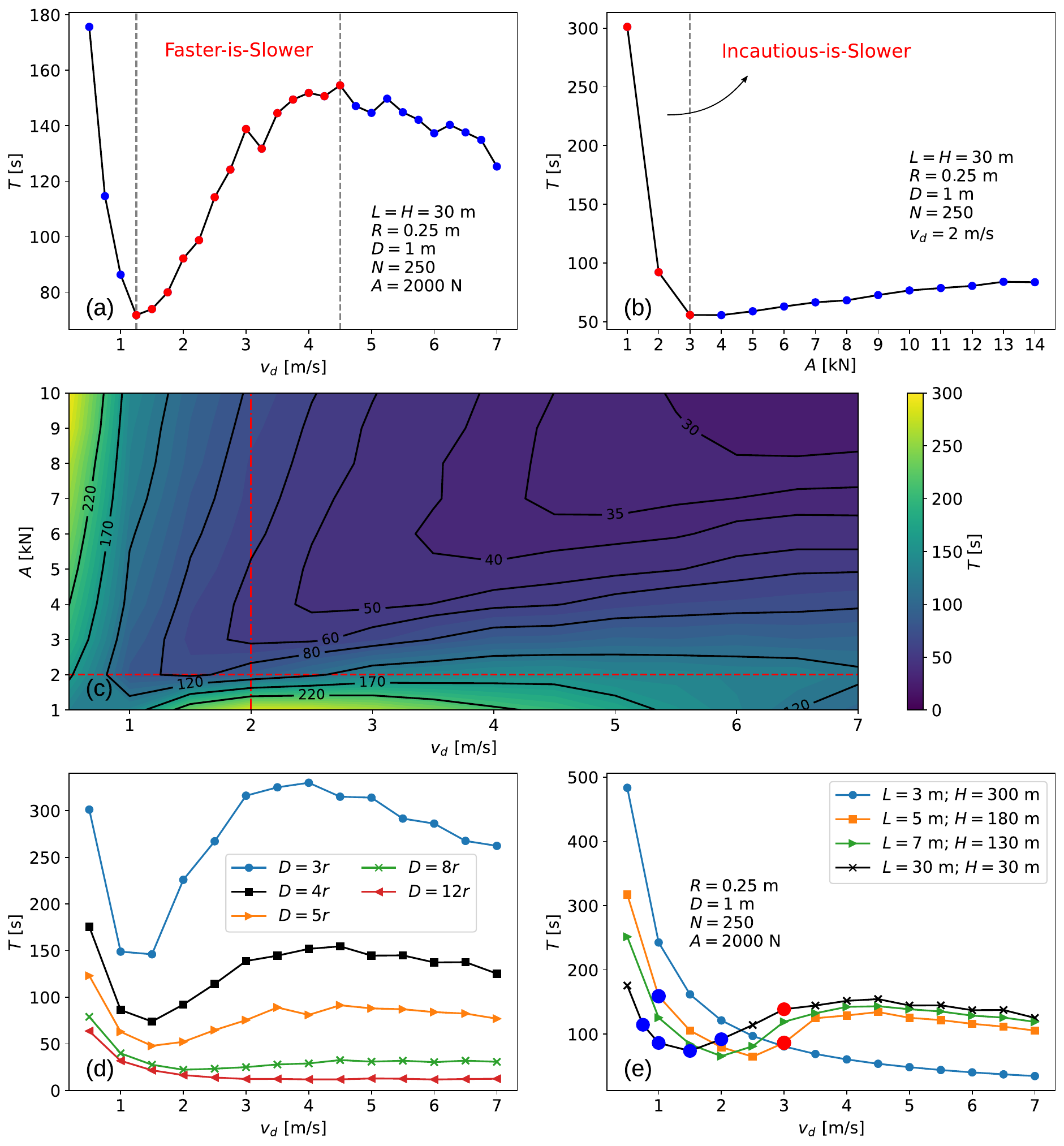}
\caption{\label{f-FIS} Evacuation time for homogeneous populations. a) Evacuation time $T$ as function of the desired velocity. b) $T$ as a function of $A$. c) Contour plot of $T$ as a function of $v_d$ and $A$. The red lines indicate the region scanned in panels a and b. d) $T$ as a function of $v_d$ for various sizes of the door. The rest of the parameters as in panel (a). e) $T$ vs. $v_d$ for different aspect ratios of the room while keeping the area $L\times H$ constant. The red and blue big dots correspond to parameter sets that will be considered for egotistic and cooperative agents throughout the work, respectively.}
\end{figure}

We begin by analyzing the case of a homogeneous crowd in a rectangular room. Although the general phenomenology for this system is known \cite{helb2000,sticco,cornes}, here we present further analysis of the dependence on the parameters and highlight certain features as a starting point for our studies. In Figure \ref{f-FIS}.a, we show the dependence of the evacuation time on the desired velocity of the agents. The typical curve exhibiting the {\em faster-is-slower} effect is obtained \cite{helb2000}. Three zones can be distinguished. First, for $v_d\lesssim 1$, we have the {\em faster-is-faster} zone in which the evacuation time decreases with the desired velocity. This corresponds to relatively ordered
evacuation processes with no relevant clogging events.  Here, the social force is stronger than the desire force. On the other hand, for $1\lesssim v_d \lesssim 4.5$, we have the faster-is-slower zone in which the evacuation time grows with $v_d$ due to the increase of clogging events that produce an intermittent flux \cite{helb2000,GARCIMARTINEXP}. This is a consequence of the fact that the desire force is stronger than the social force. Finally, for $v_d >4.5$ the evacuation time decreases with $v_d$ again, but this is a region of high pressure for which wounded and fallen pedestrians may difficult the evacuation, so additional modeling may be needed \cite{sticco,cornes-fallen}.

If, as stated in the previous section, we associate relatively small values of $v_d$ with cooperative behaviors and large values of $v_d$ with competitive or egotistic ones, the curve in Fig.\ref{f-FIS}.a indicates that neither fully egotistic nor extremely cooperative attitudes are optimal for easing the evacuation in homogeneous systems. 
In contrast, intermediate behaviors provide lower evacuation times, as discussed in \cite{victor}. The same can be said about the parameter $A$. As mentioned before, large values of $A$ can be associated with cooperative (cautious) behaviors while small values with selfish (incautious) ones, while Fig. \ref{f-FIS}.b shows that there is an intermediate value of this parameter that optimizes the evacuation process. The increase of the evacuation time for decreasing $A$ at small values of $A$ has thus a strong analogy with the faster-is-slower effect and could be named as {\em incautious-is-slower}. It is caused by the increase of clogging due to the reduction of the repulsion among agents. In general, highly competitive behaviors enhance clogging while extreme cooperative attitudes produce rather ordered but slow fluxes \cite{victor}.
 
Figure \ref{f-FIS}. c shows a contour plot of the evacuation time as a function of $v_d$ and $A$. The shape of the contours indicates that, in rather general situations, if one of the parameters is fixed, there is an optimal (non extreme) value of the other that minimizes the evacuation time. The red horizontal and vertical segments indicate the parameters scanned while plotting Figs. \ref{f-FIS}.a 
and \ref{f-FIS}.b, respectively.     

Figure \ref{f-FIS}.d shows the exit time as a function of $v_d$ considering different door widths for the same parameters as 
in Fig. \ref{f-FIS}.a. It can be seen that the positive slope of the curve in the faster-is-slower zone decreases with $D$ until the effect disappears. However, the region of $v_d$ for which the faster-is-slower effect occurs grows slightly with $D$. Throughout this work, we focus on the case $D=4 R$ with $R=0.25\,$m; nevertheless, the results in 
Fig. \ref{f-FIS}.d suggest that our main conclusions may be robust to relatively smooth changes of these 
assumptions provided the faster-is-slower effect exists. 

In Fig.\ref{f-FIS}.e, we show the exit time as a function of $v_d$ for varying values of the room width ($L$), while keeping its area and the initial number of agents constant. As $L$ decreases, the curve shifts to the right, indicating that ordered fluxes are obtained at larger velocities and higher pedestrian speeds are needed to observe the faster-is-slower regime. This is essentially due to two related effects. First, for small $L$, the rooms are narrow but large, so pedestrians coming from the back of the room require a large time to reach the door and contribute to the clogging around the exit. Second, the flux in narrow rooms is better directed towards the exit, and the angles of the collisions between agents may tend to be smaller. We will further discuss the role of the geometry of the room when analyzing the flux of mixed populations.

\subsection{Results for mixed populations with imitation of cooperative behavior} 

Here we analyze the dynamics of mixed populations in which competitive agents adopt cooperative behavior when they are close enough to a cooperative agent, as explained in the model section. We consider a crowd consisting of $N_e=250$ competitive pedestrians and a variable number $N_c$ of cooperative agents. The main objective is to investigate if the addition of cooperative agents to a competitive crowd can reduce the exit time of the original crowd. As mentioned, cooperative and egotistic agents will be distinguished by their values of the parameters $v_d$ or $A$, which will be referred to as $v_{d,c}$ (or $A_{c}$) for cooperative agents and $v_{d,e}$ (or $A_{e}$) for egotistic agents.

\begin{figure}[ht]
\includegraphics[width=0.85\columnwidth]{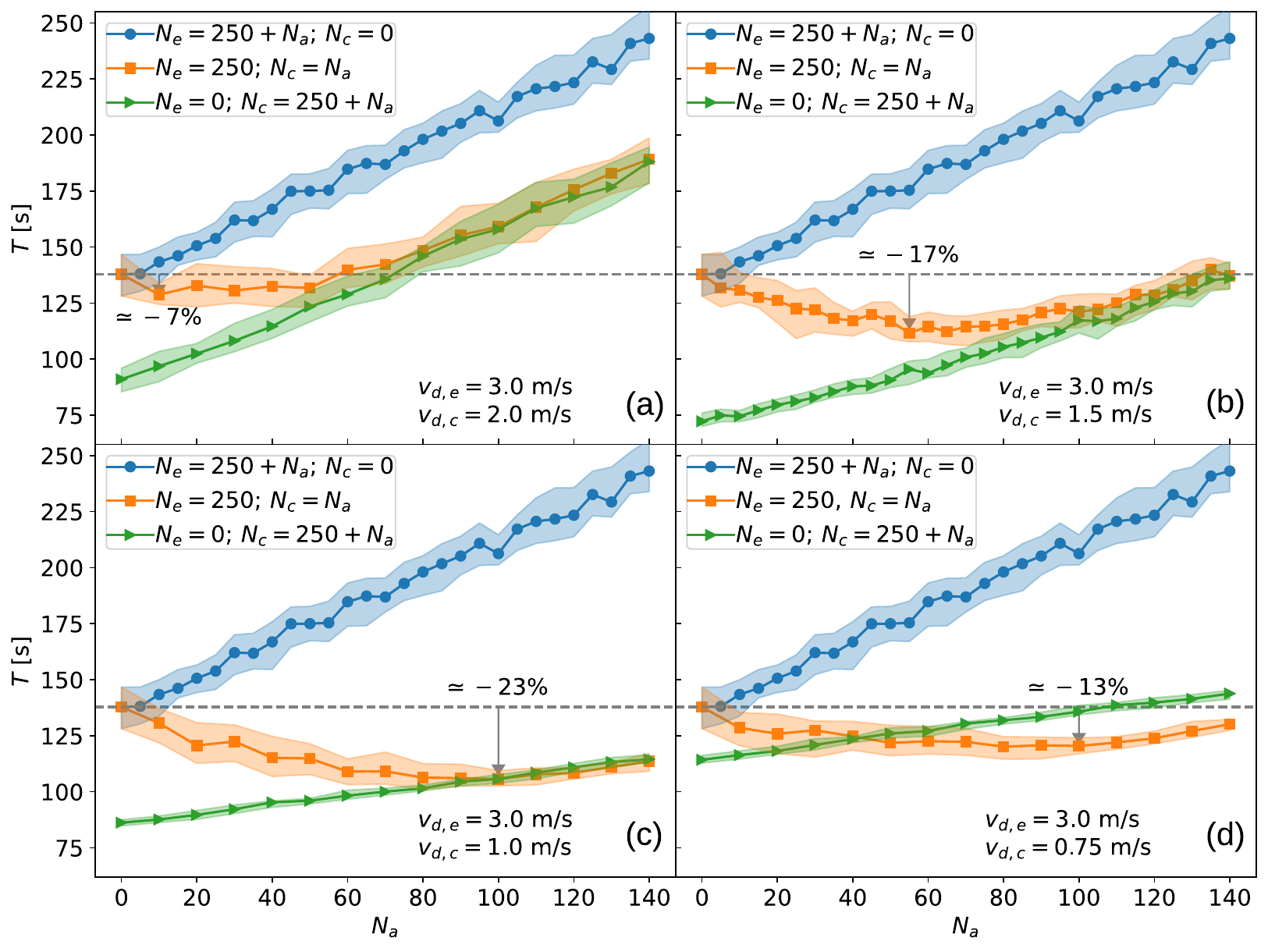}
\caption{\label{f-mix-v} Exit time for mixes with imitation in a square room with $L=H=30$m and $D=1$m.
Mixes with cooperative and egotistic agents differing in the values of $v_d$ a) Evacuation time (median) as a function of the number of added agents ($N_a$) for the populations with the parameter indicated in the panel. The blue and green curves correspond to pure egotistic and pure cooperative populations, respectively, while the orange curve corresponds to mixed populations. The colored zones represent $50\%$ confidence intervals (limited by the first and third quartiles obtained 
from $50$ realizations for each parameter set). Panels b, c, and d are the same as panel a for different values of the desired velocity of the cooperative agents.  The rest of the parameters are the ones indicated in Table \ref{tabla}.}
\end{figure}

We begin by analyzing the case of {\em patient} cooperative pedestrians ($v_{d,c}<v_{d,e}$, with $A_c=A_e$). We choose $v_{d,e}=3m/s$, so that competitive pedestrians represent hurried people which are behaving as in the faster-is-slower regime described in Fig.\ref{f-FIS}.a. Meanwhile, we consider different values $v_{d,c}$ to scan the region $v_{d,c}<v_{d,e}$ form the faster-is-slower regime to the faster-is-faster one. In Figure \ref{f-mix-v}, we show results for the evacuation time for the mixed system with imitation. Each panel corresponds to a particular value of the parameter $v_{d,c}$ and exhibits three curves for $T$ as a function of the number of added agents $N_a$. First, in orange, the curve for the mixed population of interest with fixed $N_e=250$ and varying $N_c=N_a$. Second, for the sake of comparison, we show the curve for the pure system of competitive agents with varying $N_e=250+N_a$ (with $N_c=0$), and third, the curve for the pure system of cooperative agents with $N_c=250+N_a$ ($N_e=0$). In addition, the horizontal dashed line indicates the exit time obtained for a population of $N_e=250$ with $N_c=0$.

Note that, while the blue curve for pure competitive agents is the same in all the panels, the green curve for cooperative populations sinks with decreasing $v_{d,c}$ from panels a to c, and then rises again in panel d for the smallest value of $v_{d,c}$ analyzed. This could be expected taking into account the profile of the curve for the exit time vs. $v_d$ shown in Fig. \ref{f-FIS}.a. 

To analyze the results for the mixed system, we begin by focusing on Fig. \ref{f-mix-v}.a which considers the largest
value of $v_{d,c}$ studied, so that the difference between cooperative and egotistic agents is the smallest one. 
Starting from $N_c=N_a=0$, as the number of added cooperative agents grows, the evacuation time for the mixed system
decreases to values smaller than that for to the pure competitive population of $N_e=250$ (horizontal dashed line).
Then, for $N_a\sim 50$ it begins to grow and surpasses the $N_a=0$ level. At large $N_a$ the curve for the mixed system 
approaches that for the pure cooperative population. This latter fact could be expected since a large concentration of cooperators would make all the competitive agents to behave as cooperative ones. Hence, it is worth remarking that, with imitation, a population of $N_e=250$ with a number $N_c=N_a\lesssim 50$ of added cooperative agents is evacuated faster than the pure competitive population with $N_e=250$ and $N_c=0$. The addition of cooperative agents reduces the evacuation time despite the increase in the total number of pedestrians. The maximal reduction obtained in the results of Fig. \ref{f-mix-v}.a is of the order of $7\%$ at $N_a\sim 30$. The effect is even more notable for smaller values of $v_{d,c}$ as those considered in Fig. \ref{f-mix-v}.b and Fig. \ref{f-mix-v}.c, where we observe maximal reductions of the order of $17\%$ and $23\%$ (indicated with vertical arrows), respectively. As an example of a simulation with the parameters as in Fig.\ref{f-mix-v}.c considering $N_c=75$ see video 1 in \cite{supmatamir}.  

\begin{figure}[ht]
\includegraphics[width=0.85\columnwidth]{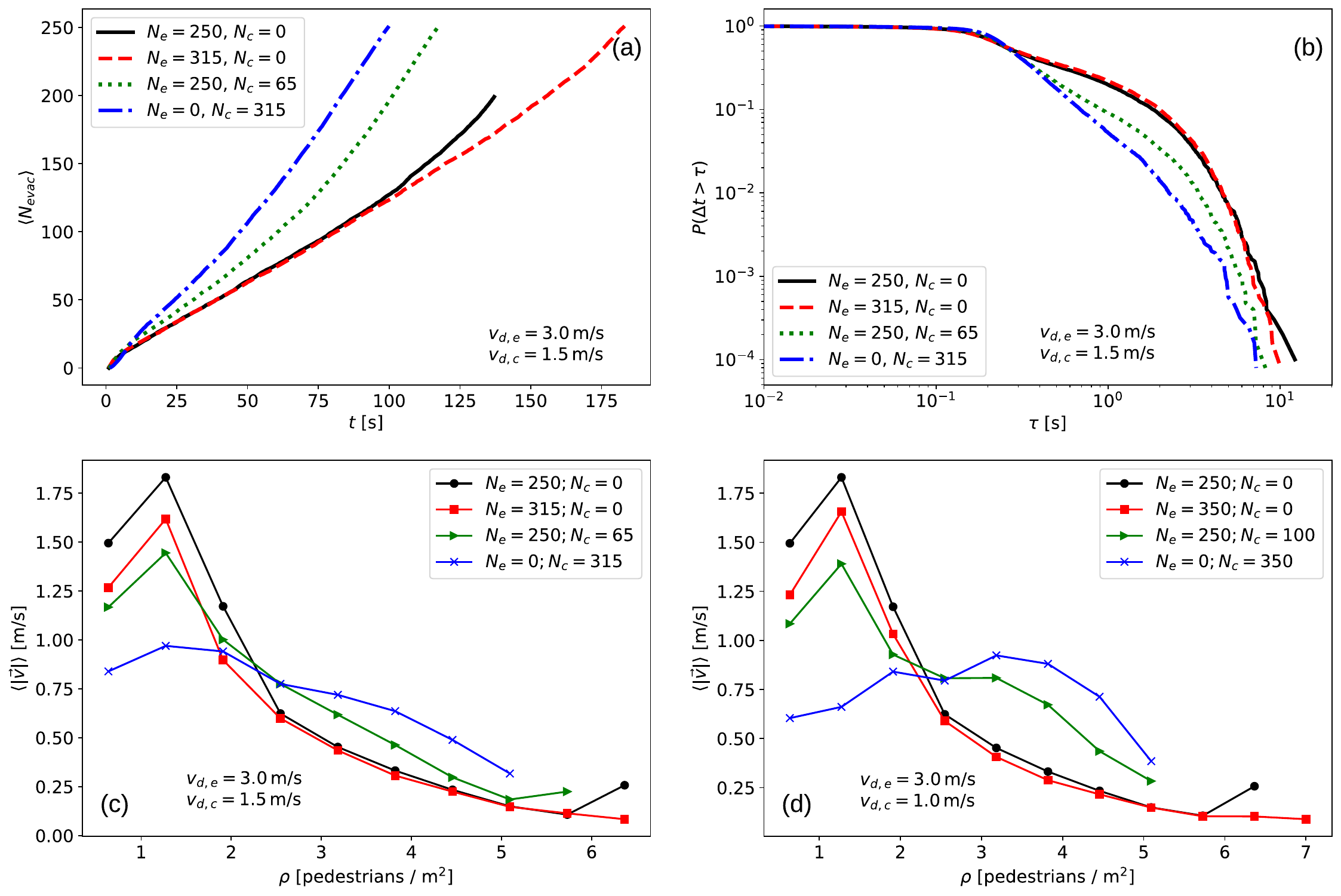}
\caption{\label{f-surv-cuadrada} a) Number of evacuated pedestrians as a function of time for various systems selected from those analyzed in Fig. \ref{f-mix-v}.b. b) Survival functions for the systems studied in panel a. c) Fundamental diagrams for selected systems analyzed in Fig.\ref{f-mix-v}.b.
(cooperative agents with $v_{d,c}=1.5$). d) Fundamental diagrams for selected systems analyzed in Fig.\ref{f-mix-v}.c (cooperative agents 
with $v_{d,c}=1$).}  
\end{figure}

Figure \ref{f-surv-cuadrada}.a shows the time dependence of the number of escaped agents for the systems studied in Fig. \ref{f-mix-v}.b
for the case $N_a=65$ (in the region of maximal reduction). It can be seen that, after a short transient, the slope of the curve corresponding to the mixed system, i.e. the outflow flux, becomes considerably larger than that for the pure competitive one. This enhancement of the flux is naturally related to a decrease in the times between successive escapes, as shown in Fig. \ref{f-surv-cuadrada}.b where we plot the survival functions for the same systems. Clearly, the presence of cooperative pedestrians reduces clogging. The short transient regime before mentioned is related to the formation of a semicircular bulk of high density of pedestrians around the door (see Fig. \ref{f-esquema}.b and video 1 in \cite{supmatamir}). Theeffects of clogging become relevant only once such a structure is established. The flat part of the curves in Fig. \ref{f-surv-cuadrada}.b observed for $\tau\lesssim 0.1$ corresponds to {\em free} flow of pedestrians while the decreasing regime corresponds to clogs.  
It is worth noting that the survival functions for the pure egoistic systems with $N_e=250$ and $N_e=315$ are nearly the same, revealing the fact that the clogging dynamics is independent of the number of pedestrians for large enough $N$. This resembles the implication of the Beverloo law in vertical arrangements of granular media \cite{beverloo}. In Figs. \ref{f-surv-cuadrada}.c and \ref{f-surv-cuadrada}.d,  we show the fundamental diagrams \cite{funddiag} for particular systems taken from Figures \ref{f-mix-v}.b and \ref{f-mix-v}.c, respectively. The values of $N_c$ for mixed systems were chosen close to those producing the maximal reduction of $T$ in each case. For densities in the range of $2\lesssim \rho \lesssim 5$, the mixed systems show a considerable enhancement of the mean velocities compared to the pure competitive populations. The gains are larger for the system with the smallest value of the desired velocity of the cooperators considered (panel d). We can clearly link the enhancement of the velocity in the exit zone caused by mixing with the previously found decrease of clogging (panel b), augmentation of flux (panel a), and decrease of evacuation time (Fig. \ref{f-mix-v}). Interestingly, the results for mean velocity vs. density for the pure competitive systems are nearly independent of $N_e$. This  supports what was said in connection with the Beverloo law. 

\begin{figure}[ht]
\includegraphics[width=0.65\columnwidth]{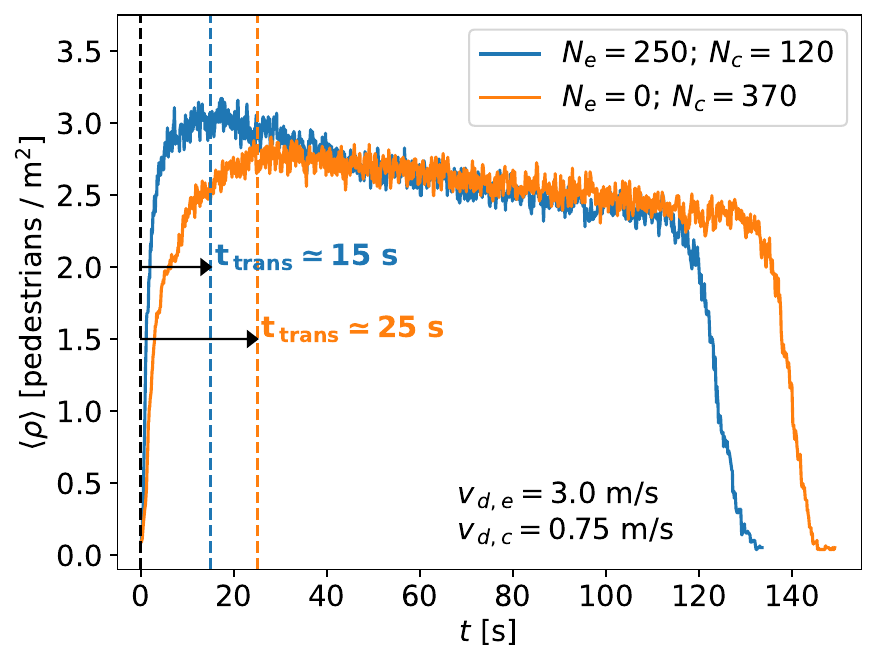}
\caption{\label{f-dens120} Density of pedestrians near the exit as a function of time. Results for systems selected from Fig. \ref{f-mix-v}.d.}
\end{figure}

The behavior of the curves for the smallest value of $v_{d,c}$ studied (Fig. \ref{f-mix-v}.d) is noteworthy. In this case, for large enough $N_a$, the curve for the mixed system is below that for the cooperative populations. This means that the evacuation time for the mixed system is not only smaller than for the pure competitive system with $N_e=250$, but also smaller than for the cooperative population with $N_c=250+N_a$. This is mainly because, for such slow cooperative agents, the initial transient regime is considerably larger for the pure cooperative system than for the mixed one. This effect can be observed in Fig. \ref{f-dens120}, where the evolution of the density of pedestrians around the door $\rho(t)$ is compared for a pure cooperative system and a mixed system with the same total number of agents. The egotistic pedestrians reach the door much faster than the cooperative ones, and the high-density regime is formed more quickly in the mixed system than in the purely cooperative system. Meanwhile, the dynamics of both high-density regimes are rather similar since a large fraction of egotistic agents behave cooperatively in the mixed system. From another point of view, it is interesting to note that when comparing Fig. \ref{f-mix-v}.c with Fig. \ref{f-mix-v}.d, the shift of the curve for the pure cooperative populations is much larger than that of the curve for mixed systems. Clearly, the decrease in $v_{d,c}$ affects a pure cooperative system more strongly than a mixed system with the same total number of agents, as could be expected. Despite the particular ordering of the curves in Fig. \ref{f-mix-v}.d, at very large $N_a$, all the egotistic agents are expected to imitate the cooperative behavior and the curve of the mixed system would approach the pure cooperative one from below.

\begin{figure}[ht]
\includegraphics[width=0.85\columnwidth]{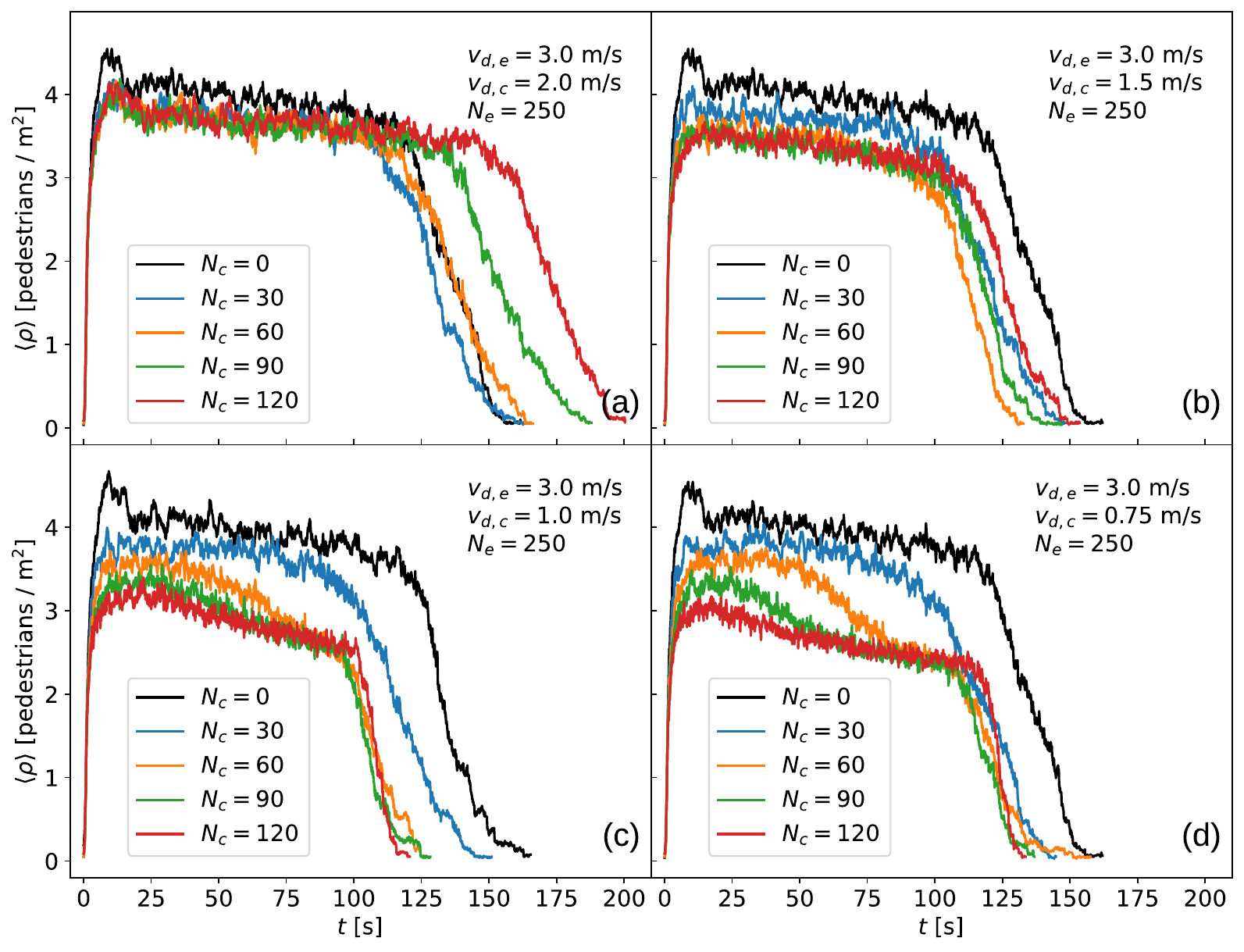}
\caption{\label{f-mix-v-dens} Density of pedestrians near the exit as a function of time. Results for mixes with imitation. Panels a, b, c
and d are in correspondence with those of Fig. \ref{f-mix-v}. }
\end{figure}

In Fig. \ref{f-mix-v-dens}, the evolution of the density of pedestrians near the exit for mixed populations is shown. The value of the parameter $v_{d,c}$ diminishes from Fig. \ref{f-mix-v-dens}.a to \ref{f-mix-v-dens}.d, in correspondence with the values considered in the panels of Fig. \ref{f-mix-v}. Each panel analyzes various values of $N_c$, including the case of $N_c=0$. Let's focus on Figure \ref{f-mix-v-dens}.a, which considers the largest value of $v_{d,c}$ studied. The results allow us to observe the change in the duration of the evacuation as $N_c$ is varied, which is in agreement with the results in Fig. \ref{f-mix-v}.a. Moreover, we can see that the addition of cooperative agents to the competitive population produces a decrease in the values of density observed along the evacuation. This decrease is on the order of $10\%$ for the peak observed at the beginning of the evolution ($t \sim 10s$). The effect is more notable for smaller values of $v_{d,c}$ (panels b, c, and d), where we see a progressive splitting of the curves for the different values of $N_c$, and a reduction of the density on the order of $25\%$ or even larger in some cases. This indicates that the addition of cooperative pedestrians with imitation not only speeds up the evacuation but also causes a decrease in the density near the exit, meaning a smaller pressure for the pedestrians and a lowering of the risk of accidents.

\begin{figure}[ht]
\includegraphics[width=0.95\columnwidth]{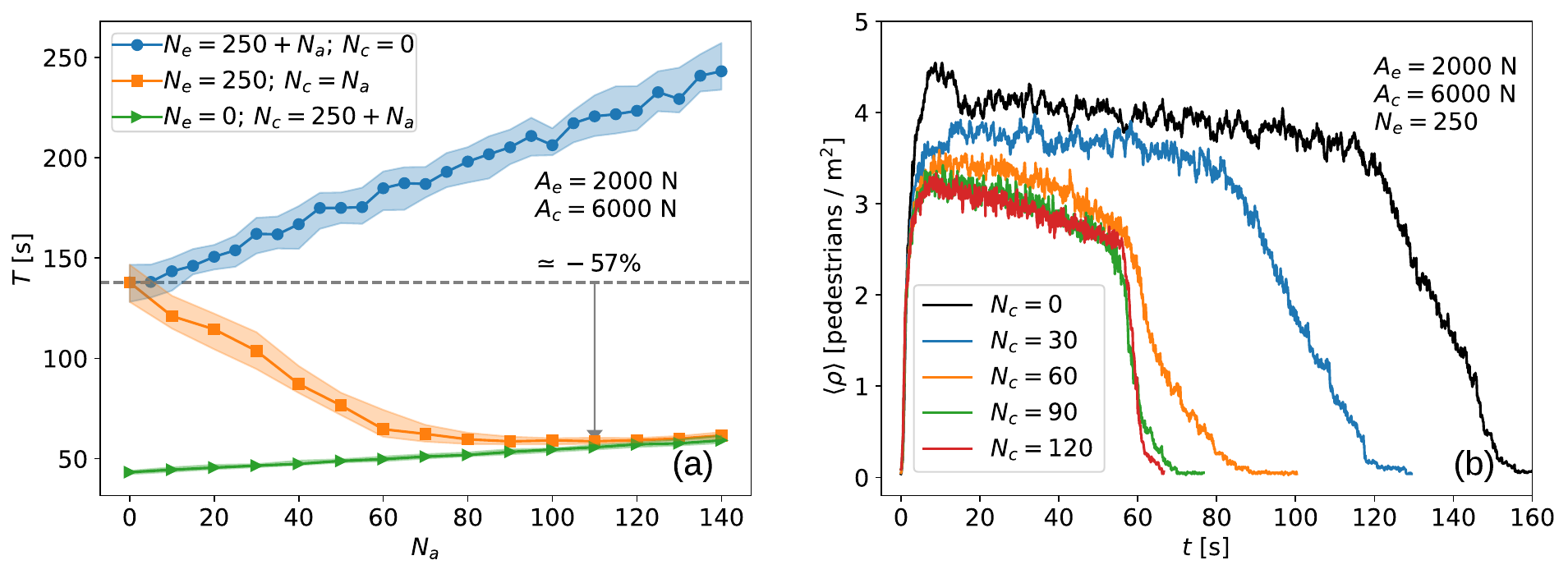}
\caption{\label{f-mix-A} Mixed systems with imitation in a square room with $L=H=30$m and $D=1$m. Cooperative and egotistic agents differ in the values of $A$ (cautious cooperative agents). a) Exit time (median) plotted as a function of the number of added agents ($N_a$) for different cases presented as in Fig.\ref{f-mix-v}. The colored zones represent $50\%$ confidence intervals as in Fig. \ref{f-mix-v}. b) Density of pedestrians near the exit as a function of time for some selected cases. The rest of the parameters are the ones indicated in Table \ref{tabla}.}
\end{figure}

We now turn to the case of cooperative agents defined through the value of $A$, i.e. what we have called cautious cooperative agents. We consider $A_{c} > A_{e}$ with $v_{d,c} = v_{d,e}$. In Fig. \ref{f-mix-A}.a, we show the results for the exit time, which are analogous to those obtained in Figs. \ref{f-mix-v}.a - \ref{f-mix-v}.c. The reduction of the exit time for the mixed system with respect to the pure competitive system with $N_e=250$ is even larger (approximately 57\%). However, here we are considering a relatively large value of $A_c$ ($A_c = 6000\,$N$\,= 3 A_e$). For smaller values of $A_c$ in the range of $2500$ to $5000$, the reduction of the exit time is of the same order of magnitude as those found in Fig. \ref{f-mix-v} (results not shown). The particularly large value of $A_c$ used in Fig. \ref{f-mix-A}.a was selected in the search for behavior similar to that found in Fig. \ref{f-mix-v}.d, where the curve for the mixed system surpasses that for the pure cooperative one. However, such behavior was not found even for values of $A_c$ as large as $10000\,$N. This supports the argument given before, which suggests that this particular phenomenon was essentially due to the small value of the velocity of the cooperators. Figure \ref{f-mix-A}.b shows the evolution of the density near the exit for the mixed system for various values of $N_c$, considering the rest of the parameters as in Fig. \ref{f-mix-A}.a. The results are analogous to those found in Fig. \ref{f-mix-v-dens}. Again, the presence of cooperators decreases not only the exit time but also the density of pedestrians near the door.

\begin{figure}[ht]
\includegraphics[width=0.95\columnwidth]{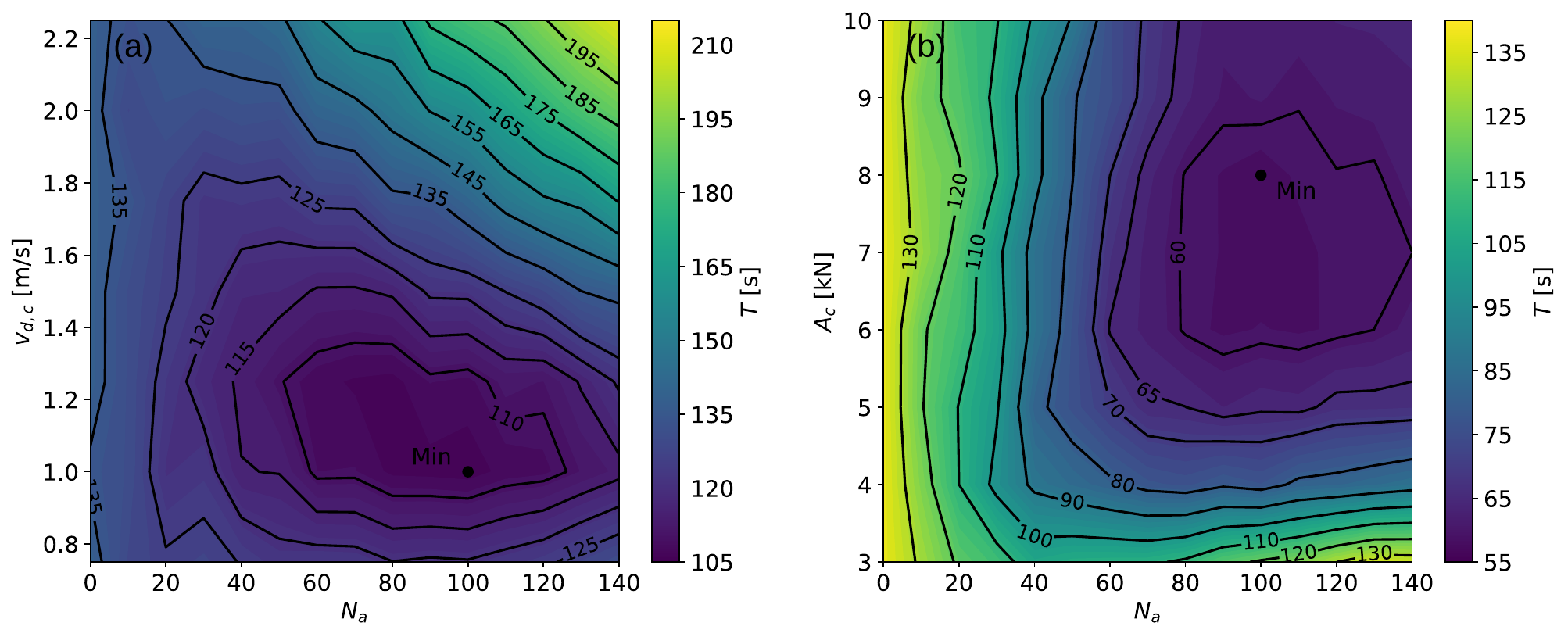}
\caption{\label{f-contours} Contours for the exit time of mixed systems. a) Exit time as a function of $N_c$ and $v_{d,c}$ 
for mixed systems with cooperative behavior defined by the value of $v_{d,c}$. The parameters are the same as those 
in Fig. \ref{f-mix-v}, with $N_e=250$. b) Exit time as a function of $N_c$ and $A_{c}$ 
for mixed systems with cooperative behavior defined by the value of $A_{c}$. The parameters are the same as
those in Fig. \ref{f-mix-A}, with $N_e=250$.}
\end{figure}

In Figure \ref{f-contours}, we summarize the results for the exit time of the systems with imitation of cooperative behavior analyzed in this section. Fig. \ref{f-contours}.a considers the case of cooperators defined through $v_d$. A contour plot of the exit time as a function of $N_c$ and $v_{d,c}$ is shown. Meanwhile, Fig. \ref{f-contours}.b shows the exit time as a function of $N_c$ and $A_c$ for the case of cooperators defined through the latter parameter. In both cases, we find a point of minimal exit time. Additionally, regions of relatively low exit time with not very large values of $N_c$ can be clearly identified.

\subsection{The case of long rooms} 

Here, we focus on evacuations of long rooms or corridors, where, as we will see, the dynamics may result differently than for a square room. We consider a {\em long} rectangular room with dimensions $L=5\,$m and $H=180\,$m, which have the same area as the square room considered in the previous sections.

\begin{figure}[ht]
\includegraphics[width=0.95\columnwidth]{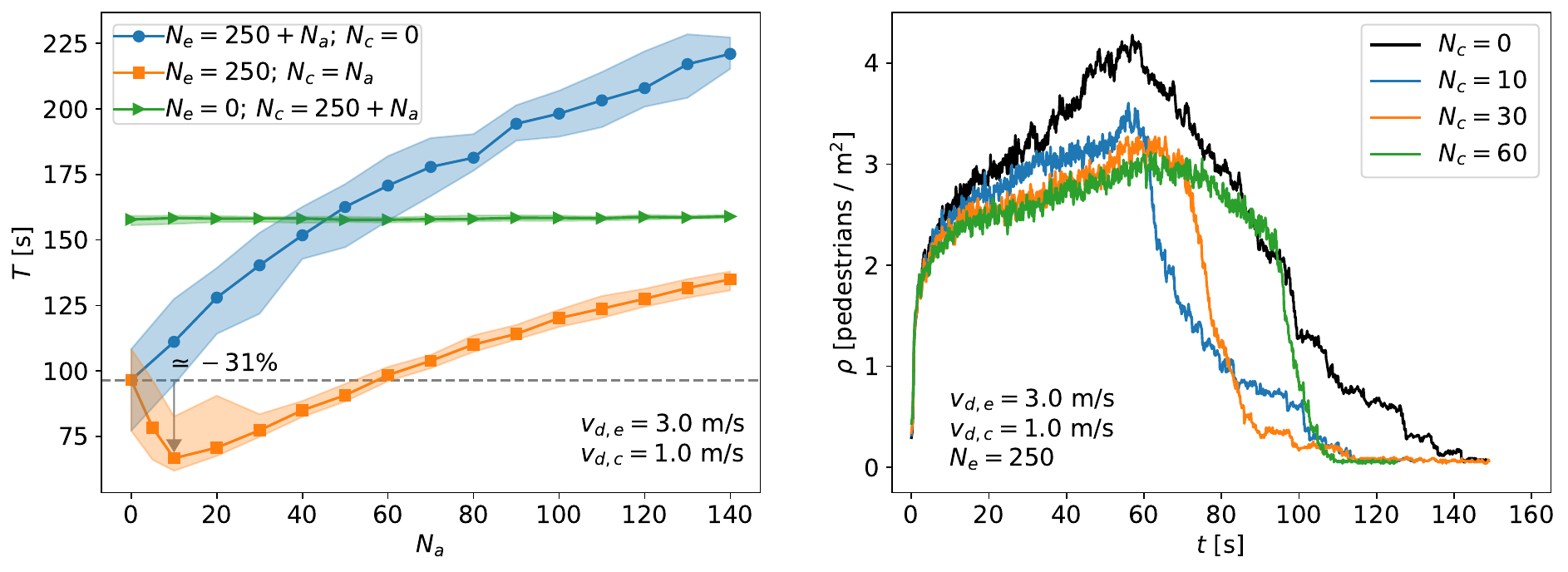}
\caption{\label{f-pasillo} Mixed systems with imitation in a {\em long} room with $L=5m, H=180$m and $D=1$m.
Cooperative and egotistic agents differ in the values of $v_d$. a) Exit time as a function of the number of added
agents for the cases indicated in the panel. The colored zones represent $50\%$ confidence intervals as in Fig. \ref{f-mix-v}. b) Density near the exit as a function of time for selected systems
from panel a.}
\end{figure}

First, we study the dynamics of evacuation for pure cooperative, pure egotistic, and mixed populations with the same parameters as those considered in Fig. \ref{f-mix-v}.c. so that the only thing that changes is the aspect ratio of the room. Figure \ref{f-pasillo}.a shows the evacuation time 
for the three types of populations as a function of the number of agents added. It is worth noting that, in contrast to what happens for the square room, here the evacuation time for the pure cooperative population with $N_c=250$ ($N_e=0$) is larger than that for the pure competitive one with $N_e=250$ ($N_c=0$). This can also be seen in Fig. \ref{f-FIS}.e by comparing the blue and red spots denoting the parameters for cooperative and egotistic agents, respectively. (See the spots at $v_d=1 m/s$ and $v_d=3 m/s$ for different values of $L$.) In Fig. \ref{f-pasillo}.a it is also notable that the evacuation time for the cooperative population is independent of $N_c$ within the range studied. This latter effect is due to the fact that, in such a long room, the cooperators are in the faster-is-faster regime. Hence, they perform an essentially free walk with almost no physical contact and no clogs. The evacuation time is just the time it takes an agent located at the back of the room to reach the exit walking at $v_{d,c}$. However, the situation may change for a larger number of agents since, as the initial density increases, clogs would begin to occur and the evacuation time should grow. An interesting point of the results in \ref{f-pasillo}.a is that, although the cooperators have a larger evacuation time than the egotistic agents, the addition of a small number of cooperators to the competitive population reduces considerably the effective evacuation time of the latter. The maximal reduction, of more than $30\%$, is observed  for $N_a\sim 10$. This optimal value of $N_a$ is remarkably smaller than those found in the square room, and actually rather small compared to $N_e=250$. 

A noteworthy fact observed in the simulations (see video 2 in \cite{supmatamir}) is that, due to the small width of the room, single isolated cooperative agents induce the formation of slow moving clusters composed of imitators. Such clusters partially block the advance of fast egotistic pedestrians, decreasing their effective flux towards the exit and reducing clogging at the exit. This effect helps to lower the evacuation time. However, it is important to remark that we say that clusters {\em partially} block the advance of egotistic agents because these fast agents are occasionally capable of overtaking the slow clusters and even forming fast corridors at one side of a cluster. The curves for the evolution of the density near the exit depicted in Fig. \ref{f-pasillo}.b help us to better understand all the phenomena. It can be seen that  when passing from $N_c=0$ to $N_c=10$, there is a strong decrease in clogging around the exit. Still, with such a low density of cooperators, a portion of egotistic agents manage to arrive at the door rapidly and to escape with small difficulties due to the relatively low clogging, therefore optimizing the flux. If the number of cooperators is further increased (see curves for $N_c=30, 60$ in Fig.\ref{f-pasillo}.b), clogging is reduced even more. However, in such situations, most of the competitive agents behave cooperatively and they advance slowly to the door, thus slowing down the evacuation process.
The optimal value of $N_c$ is obtained through a competence between the reduction of clogging caused by the partial blockage of the advance of the fast-moving agents and the decreasing of the number of fast-moving agents, both effects increasing with $N_c$. A last fact to note about the results in \ref{f-pasillo}.b is that the maximum of the density around the door is obtained at a much larger time that in the square room (compare to \ref{f-mix-v-dens}.c). While in the latter case, the initial rapid growth of the density (lasting for about $10$ seconds) gets the system to the maximal density and is then followed by a slow decay, in the long room, the initial rapid growth is not that fast, and after it, the density continues to grow slowly until reaching a maximum at about $60\,$s. Such a difference between the density profiles of the square room and those of the long room occurs even for the case $N_c=0$. Therefore, it should be mainly caused by the geometrical characteristics of the room. Note that in the long room, the growth of the density around the door is due to the quasi one-dimensional flow of agents that approach the exit essentially from a single direction. In contrast, in the square room, pedestrians arrive at the door from all directions (with {\em angle of incidence} ranging from $0$ to $\pi$). 

\begin{figure}[ht]
\includegraphics[width=0.95\columnwidth]{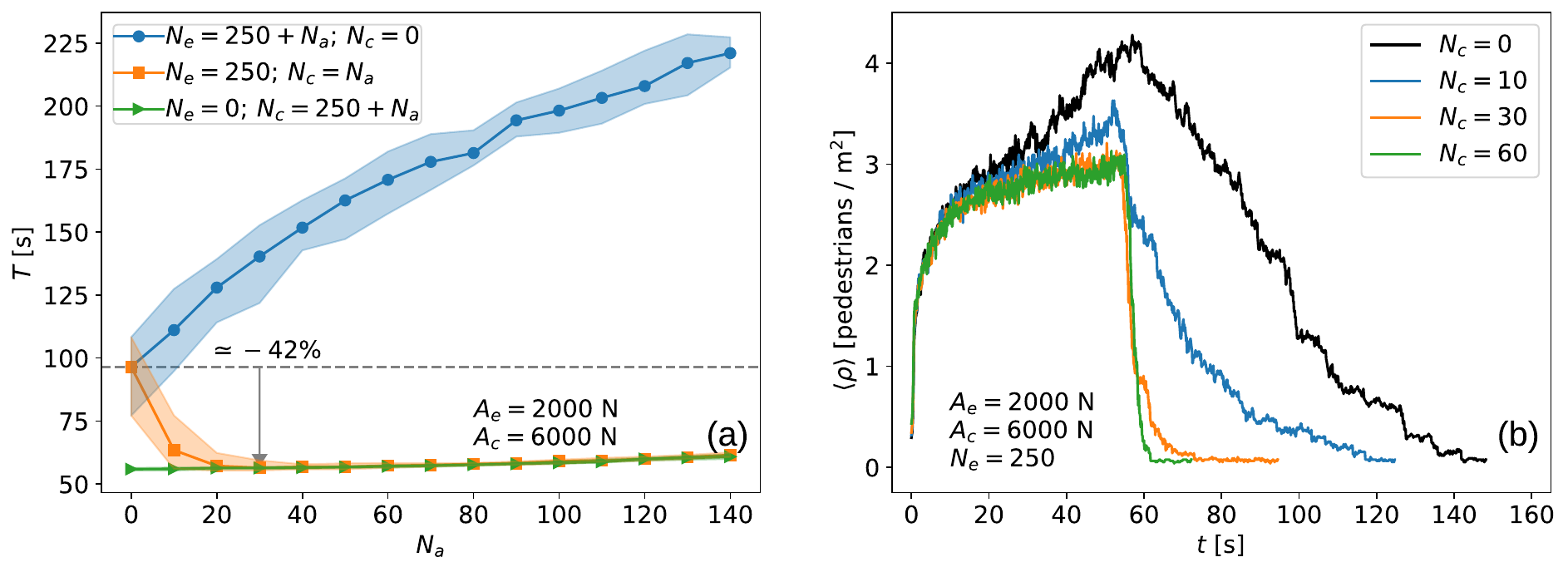}
\caption{\label{f-pasillo-A} Mixed systems with imitation in a {\em long} room with $L=5m, H=180$m and $D=1$m.
Cooperative and egotistic agents differ in their values of $A$. a) Exit time as a function of the number of added
agents for the cases indicated in the panel. The colored zones represent $50\%$ confidence intervals as in Fig. \ref{f-mix-v}.
b) Density near the exit as a function of time for selected systems
from panel a.}
\end{figure}

Now we turn to the case in which the cooperative agents are distinguished by their values of $A$. The results for the long room are shown in Fig. \ref{f-pasillo-A} and need to be compared with those in Fig. \ref{f-mix-A}, as the systems being analyzed differ only in the aspect ratio of the room. We observe that, similar to the square room, the exit time for pure cooperators in the long room is smaller than that for pure competitive agents. However, the two main effects of changing the geometry on the results for mixed populations are the same. Specifically, the optimal value of $N_c$ in the long room (Fig. \ref{f-pasillo-A}.a) is much smaller than in the square room (Fig. \ref{f-mix-A}.a), and the maxima of the density profiles are achieved at much larger times compared to the square room (compare Fig. \ref{f-pasillo-A}.a to Fig. \ref{f-mix-A}.b).
It should be mentioned, however, that in contrast to the case analyzed in Fig. \ref{f-pasillo}, when cooperative agents are defined by the value of $A$, there is no formation of slow-moving clusters, and the reduction in evacuation time is due only to the lower density observed in the exit zone.

\subsection{Mixed populations with combined cooperative behaviors}

Up to now, we have considered separately the cases of patient (small $v_d$) and cautious (large $A$) cooperative agents since each one is related to a particular model parameter. However, it is reasonable to think that cooperative pedestrians in real life would combine patience and cautiousness in diverse manners. To check the consistency of our results, we have performed two additional types of simulations in which these prototypical cooperative attitudes appear combined in different ways. On the one hand, we considered simulations of mixed populations that include the three types of agents together, i.e.  competitive agents, patient cooperative agents and cautious cooperative agents. On the other hand, we performed simulations of populations with a single type of cooperative agents but considering that each cooperative agent has both characteristic together (small $v_d$ and large $A$).

Excepting for the imitation mechanism, the assumptions for the three species dynamics are the same as those for two species.
In order to define the imitation mechanism for the three species system we consider a majority game as follows. Lets assume that at a given time there are $n_1$ patient cooperative pedestrians and $n_2$ cautious cooperative pedestrians inside the imitation radius of a certain competitive agent. Then, if $n_1>n_2$ ($n_2>n_1$) the competitive pedestrian adopts the parameters of the patient (cautious) agents. Meanwhile, if $n_1=n_2>0$ there are two possibilities. First, if before the equality begins to hold, the competitive agent was imitating a given type of cooperative pedestrian, then it persists imitating the same type. Second, if the competitive agent was not imitating, it chooses at random the cooperative behavior to imitate. For instance, if at a given time we have $n_1=1$ and $n_2=0$ (thus the competitive agent is behaving as patient) and suddenly a cautious agent enters the circle so that we turn to have $n_1=n_2=1$, then the competitive agent continues behaving as patient. On the other hand, if the competitive agent is not imitating because we have $n_1=n_2=0$ and, suddenly, two cooperative agents of different type enter the circle at the same time step, the competitive agent chooses at random between the two cooperative behaviors. We remark that a competitive pedestrian behaves as competitive at any time for which $n_1=n_2=0$ holds. 

\begin{figure}[ht]
\includegraphics[width=0.95\columnwidth]{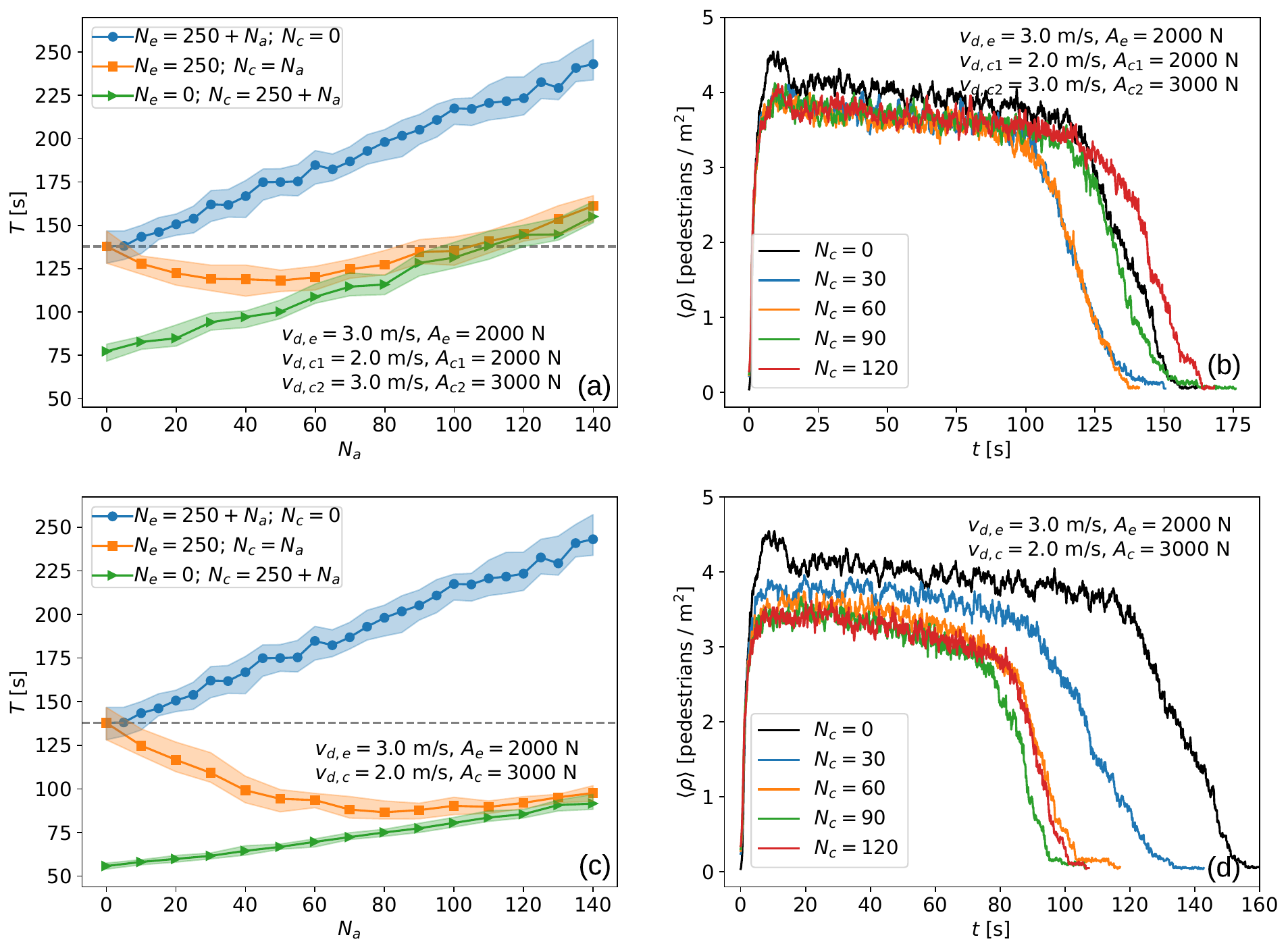}
\caption{\label{f-combined1} Mixes with combined cooperative behaviors in a square room. a) Exit time as a function of the number of added agents for the cases indicated in the panel. In this case, the mixed system is a three species system, as explained in the text. b) Density near the exit as a function of time for selected systems
from panel a. c) Exit time as a function of the number of added agents for the cases indicated in the panel. In this case, the mixed system is a two species system in which all the cooperative agents are both patient and cautious. d) Density near the exit as a function of time for selected systems from panel c. In all the calculations, the parameters are the same as those in Fig. \ref{f-mix-v}, excepting for those indicated in the panels.}
\end{figure}

For our simulations with three species we consider the same square room as in previous sections with a number $N_{c1}$ of patient cooperative agents (parameters $v_{d,c1}<v_{d,e}$, $A_{c1}=A_{e}$), $N_{c2}$ cautious cooperative agents (parameters $v_{d,c2}=v_{d,e}$, $A_{c2}>A_{e}$) and $N_e$ competitive pedestrians. In all the cases we assume $N_{c1}=N_{c2}$ and vary the total number of cooperative agents $N_{c}=N_{c2}+N_{c2}$. In Fig. \ref{f-combined1}.a we show the evacuation time for the three species system as a function of the number of added cooperative agents and compare with the results for the pure competitive systems with the same total number of agents. The results for a pure cooperative system with the two types of cooperative agents are also shown for the sake of comparison. We see that the inclusion  of a mix of the two types of cooperative agents into the competitive population leads to the same type of behavior observed in systems for which all the cooperative agents are of the same type (compare with Figs. \ref{f-mix-v} for patient agents and Fig. \ref{f-mix-A} for cautious ones). Figure \ref{f-combined1}.b shows the density near the exit as a function of time for various values of the total number of cooperators. Again, as in Figs. \ref{f-mix-v-dens} and Fig. \ref{f-mix-A}.b we find that the presence of cooperators reduces the density near the exit. Note that the minimum in the exit time for the three species system found in Fig. \ref{f-combined1}.a is not as deep as in other previously shown examples, and the reduction of density is not so notable. This is because we have chosen a moderate degree of cooperation for both types of cooperative agents, as done for instance in Fig. \ref{f-mix-v}.a for the case of patient cooperative behavior. Further decreasing of $v_{d,c1}$ or increasing of $A_{c2}$ would lead to more notable effects. 

The reason why we have chosen such relatively small degrees of {\em patience} and {\em cautiousness} for the cooperators in the previous example is that we want to compare with the case in which all the cooperators are at the same time patient and cautious, with the same parameters. For this we consider a system with a single type of cooperators with parameters $v_{d,c2}<v_{d,e}$ and $A_{c2}>A_{e}$. The imitation mechanism in this case is the same as that considered before for a single species of cooperative agents, but now, when a competitive pedestrian imitates a cooperative agent, it modifies both its value of $v_d$ and $A$. The results for the evacuation time and the density near the exit for this system are shown in Fig. \ref{f-combined1}.c and Fig. \ref{f-combined1}.d, respectively. As expected, the effects of reduction of $T$ and decreasing of the density are enhanced with respect to the example analyzed in Figs. \ref{f-combined1} a and Fig. \ref{f-combined1}. b, because now all the cooperative agents share the two characteristics. However, the main point here is that, by analyzing two different forms of combining the two types of cooperative behaviors considered, the results in this section show the consistency and robustness of our previous conclusions regarding the effect of imitation of cooperation.

\subsection{Mixed populations with no imitation}

The results in previous sections for the social force model indicate that the evacuation of a competitive crowd can be eased by the addition of cooperative agents in the case that the cooperative behavior is imitated by ordinary (competitive) pedestrians. The maximal reduction of the median of the evacuation time found was of the order $10\%$ to $50\%$, depending on the parameters and geometries. Previous studies \cite{nicolas2018} that considered agent-based and cellular-automata models for analyzing the effect of the addition of cooperative agents into a crowd of competitive ones {\em without} imitation reported much smaller reductions of the exit time, of the order of $3\%$ or smaller.

\begin{figure}[ht]
\includegraphics[width=0.95\columnwidth]{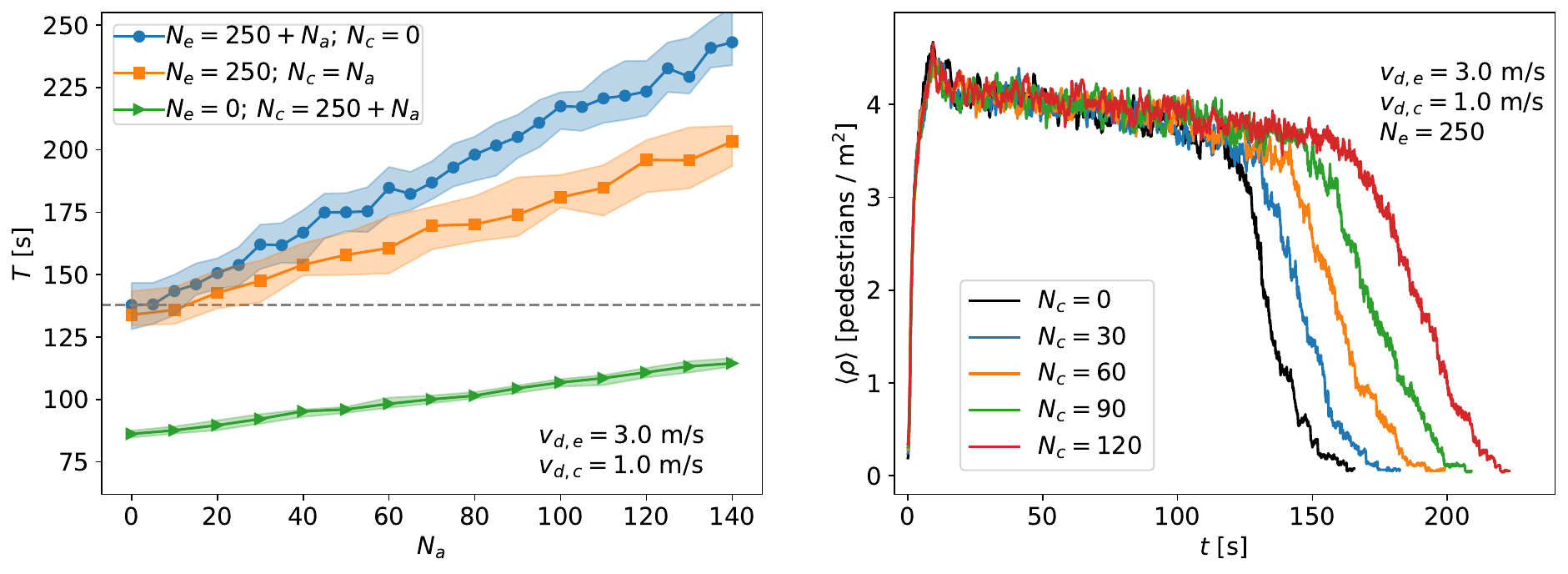}
\caption{\label{f-sincontag} Mixes with no imitation. Exit time as a function of the number of added
agents for the cases indicated in the panel. The colored zones represent $50\%$ confidence intervals as in Fig. \ref{f-mix-v}.
Parameters: $A_c=A_e=2000$N. (b) Density near the exit as a function of time for selected systems from panel a.}
\end{figure}

For the sake of comparison and completeness, here we explore the case without imitation using the social force model. 
In Fig. \ref{f-sincontag}, we show results for the exit time and density for systems with the same parameters as those used in Fig. \ref{f-mix-v}.c (and \ref{f-mix-v-dens}.c) but considering no imitation. At a first glance, the results in Fig. \ref{f-sincontag}.a indicate that the exit time for the mixed system with no imitation (orange curve) increases essentially monotonically with the number of added cooperative agents. Note however that, due to the error associated to the calculation of the median, we cannot rule out the possible existence of a tiny reduction of order $\sim 1\%$ in the median at small values of $N_a$, (i.e. a minimum in the orange curve at a non vanishing value of $N_a$). This would be in agreement with the findings in \cite{nicolas2018}. Unfortunately, to verify this would require a very large number of simulations ($\sim 5000$ realizations instead of $50$ to reduce the error of the median in one order of magnitude). We find this unnecessary because the effect, if existent, has limited statistically significant since it would correspond to a decreasing of $\sim 1\%$ for the median while the iterquartile distance (the width of the orange shaded area) is of the order of $\sim 10\%$. 

Still, we have some relevant things to mention. Regarding the results in Fig. \ref{f-sincontag}.a, it is important to emphasize that, taking the system with $N_e=250$ and $N_c=0$ as starting point, the addition of cooperative agents results in a much slower growth of the exit time compared to the addition of competitive pedestrians. In other words, at a constant total number of agents, it is always advantageous to have a fraction of them behaving as cooperative. This is in agreement with the findings in \cite{nicolas2018}. Regarding Fig. \ref{f-sincontag}.b
we see that the density in the exit zone results essentially independent of the number of added agents (compare to Fig. \ref{f-mix-v-dens}.c), except at the end of the evacuation. Hence, no reduction of the pressure felt by the agents is expected. Again, the results are noisy and much more detailed calculations would be needed in order to determine whether there is a small reduction of density or not for a small for small values of $N_c$, but in any case the effect would be of very limited significance compared to that found in the cases with imitation. 

As part of our studies, we have also analyzed mixed systems with no imitation with the cooperative behavior defined by the parameter $A$ (with the parameters as in Fig. \ref{f-mix-A}). The obtained curves (not shown) are analogous to those in Fig. \ref{f-sincontag}. The conclusions are the same as for systems with cooperation defined through the parameter $v_d$. Moreover, we have studied mixed systems with no imitation in long rooms with parameters as in Figs. \ref{f-pasillo} and \ref{f-pasillo-A}, and arrived at the same main conclusion:  the addition of cooperative agents that are not imitated by competitive pedestrians does not reduce significantly the exit time $T$ of the original crowd within the social force model, but it produces a slower growth of $T$ compared to adding competitive pedestrians.

\section{Final remarks and conclusions}

By considering a standard version of the social force model, we have studied the effect of cooperation and imitation of cooperative behaviors in various scenarios of pedestrian evacuation. Our results show that the addition of a relatively small number of cooperative agents, whose behavior is imitated by the pedestrians of a crowd, can reduce the evacuation time of the crowd and the density near the exit door. This means faster and safer evacuations. The results depend on the aspect ratio of the room, with notable effects observed in long rooms such as corridors when a small number of cooperative agents are added. This highlights the influence of the room shape and geometries, especially in the presence of populations with inhomogeneous behaviors. 

It is important to stress the fact that the reduction of the evacuation time obtained by adding cooperative agents in the system with imitation occurs despite the total number of agents being larger. This means, more pedestrians evacuate faster if  some of them are cooperative. In the absence of imitation, the reduction is not significant within our model. 
However, the addition of cooperative agents produces an increase that is smaller than that caused by the augmentation 
of the number of competitive agents. Hence, at constant total number of agents, the presence of cooperative agents is always desirable.

In most of our studies, we have considered separately the cases of patient (small $v_d$) and cautious (large $A$) cooperative agents. However, as it is reasonable to think that cooperative pedestrians in real life would combine patience and cautiousness, we have verified that different combination does not modify our main conclusions. For this we have performed simulations with three types of agents and other simulations in which cooperative agents are both patient and cautious. In all the cases we find that the addition of cooperative agents leads to faster evacuations with reduced density near the door.

Our results highlight the relevance of cooperative attitudes in facilitating evacuations, particularly emphasizing the advantages of achieving the imitation of cooperative behaviors within a crowd. Interestingly, as mentioned in the introduction, there is evidence that cooperative attitudes emerge in dangerous or emergency situations (see Ref. \cite{CHENG2018485} and references therein). It's difficult to determine how many potential tragedies in crowds have been averted by this spontaneous mechanism. Certainly, not enough to prevent all, as tragedies occur occasionally. Hence, it is reasonable to think that educating and training people on appropriate protocols for individual behavior within crowds may help the spontaneous tendency to cooperate and may further contribute to avoiding tragedies.

A straightforward interpretation of our results suggests that people should be taught to reduce haste and enhance cautiousness in emergency situations involving crowds, by trying for instance to advance more slowly and to leave more space between pedestrians, avoiding pushing. Moreover, they should also be trained to try to persuade their neighbors to adopt similar behaviors and to remember that if neighbors are behaving collaboratively in a dangerous situation, one should also strive to do the same. While this may hold true, such a perspective could be overly simplistic or naive. Things are probably not so straightforward. Since our conclusions are limited to the predictions of the social force model, further research involving numerical simulations with other models, controlled experiments, and analysis of real-life scenarios should be conducted before making concrete decisions on education. Additionally, considerations of individual and mass psychology should be taken into account. Protocols for individual behavior within congested crowds should be designed by interdisciplinary teams. Furthermore, the recommended protocol may depend on the particular type of scenario and society. Nevertheless, even though many more studies are needed, our work aims to draw attention to the importance of designing protocols and educating people on how to behave within congested crowds. Our studies suggest that even if only a limited fraction of pedestrians remain calm and are able to induce calmness in others, the risks can be considerably reduced.


It is interesting to note that protocols and instructions for different types of emergencies usually require individuals to behave in ways that go against their instincts. For instance, during depressurization events in planes, adults responsible for children must put on their own oxygen masks before assisting the kids. Most adults accustomed to flying are aware of this recommendation and probably would follow it in an emergency, even though it seems to be rather against the natural instincts of most mothers and fathers. Similarly, people who walk in wild areas and National Parks know that, in the case of being charged by a bear, they should lie on the ground and play dead if the bear is brown, but they should act as being as big as possible and fight back in the case that the bear was black. If people can learn and follow these instructions, we believe that they could also be trained on how to behave within a high-density enclosed crowd, even if the convenient behavior may be counter to their instincts of running, pushing and escaping. 

\section{Aknowledgements}

The authors would like to thank the financial support from CNEA and CONICET, both Argentinean public institutions.

\end{document}